\documentclass[doublecol]{epl2} 

\usepackage{letltxmacro}
\usepackage{amsmath}
\usepackage{amssymb}

\newcommand{\blue}[1]{\textcolor{blue}{#1}}

\usepackage[colorlinks=true,citecolor=blue,linkcolor=blue,urlcolor=blue]{hyperref}

\title{Non-equilibrium Berezinskii-Kosterlitz-Thouless transition in driven-dissipative condensates}
\shorttitle{Non-equilibrium BKT transition in driven-dissipative condensates} 

\author{P. Comaron\inst{1,2} \and I. Carusotto\inst{3} \and M. H. Szyma\'nska\inst{4} \and N. P. Proukakis \inst{1} }
\shortauthor{P. Comaron \etal}

\institute{                    
	\inst{1} Joint Quantum Centre (JQC) Durham-Newcastle, School of Mathematics, Statistics and Physics,
	Newcastle University, Newcastle upon Tyne, NE1 7RU, United Kingdom\\
	\inst{2} Institute of Physics, Polish Academy of Sciences, Al. Lotników 32/46, 02-668 Warsaw, Poland\\
	\inst{3} INO-CNR BEC Center and Universit\`a di Trento, via Sommarive 14, I-38123 Povo, Italy\\
	\inst{4} Department of Physics and Astronomy, University College London, Gower Street, London, WC1E 6BT, United Kingdom
}

\abstract{
	We study the \textit{2d} phase transition of a driven-dissipative system of exciton-polaritons under non-resonant pumping. Stochastic calculations are used to investigate the Berezinskii-Kosterlitz-Thouless-like phase diagram for experimentally realistic parameters, with a special attention to the non-equilibrium features.
}

\begin{document}
	
	\maketitle
	
	Phase transitions are ubiquitous in nature, both within the classical and quantum realms. Dimensionality and symmetry are crucial ingredients for the determination of {the types} of phase transition (PT) {that a given} system may undergo. In a $3d$ system at thermal equilibrium, Bose particles can exhibit off-diagonal long range order (ODLRO) when driven by a control parameter below a specific critical temperature.
	This phenomenon is associated with the appearance of a Bose-Einstein Condensate (BEC), predicted to occur in {both} uniform and confined systems~\cite{pitaevskii2003bose}. In $2d$ systems, instead, the presence of thermal fluctuations destroys ODLRO, compromising the existence of a possible PT to an ordered state {at any finite temperature}~\cite{MerminWagner1966}. Nevertheless, it has been shown that {a different kind of} PT to a {quasi-condensate} state may still occur, {with the decay of correlation functions going from an exponential to a much slower algebraic law~\cite{Berezinskii1971,Berezinskii1973}. {This Berezinskii-Kosterlitz-Thouless (BKT) transition} can be pictorially understood} in terms of {the thermally activated vortices, which change their spatial distribution when crossing the critical temperature: at high-temperature they proliferate and are freely moving, at low temperatures they are much less numerous and are bound in pairs, so their detrimental impact on the coherence gets dramatically suppressed.}
	
	{The physics becomes even more intriguing when one moves away from isolated systems to driven-dissipative ones \cite{carusotto2013quantum,proukakis_snoke_littlewood_2017,hohenberg1977theory}, whose stationary state is no longer determined by thermal equilibrium, but by a non-equilibrium balance of driving and dissipation. A most celebrated platform to study this physics is based on exciton-polaritons in semiconductor microcavities, namely bosonic quasiparticles that arise from the strong coupling between light and matter excitations. These quasiparticles have a finite lifetime, which calls for some external pumping to continuously compensate for losses \cite{carusotto2013quantum}. As in standard equilibrium BEC, for sufficiently high densities a macroscopic fraction of the polariton gas condenses into a single momentum state and order develops across the whole finite sample \cite{kasprzak2006bose}. In spite of this apparent simplicity, the full characterization of the PT and of its critical fluctuations in terms of universality classes is still at the centre of an intense debate, in particular given their intrinsically $2d$ nature. A strong attention has been devoted, both experimentally \cite{hohenberg1977theory,pitaevskii2003bose,caputo2018,stepanov2019dispersion,ballarini2020directional} and theoretically \cite{Berezinskii1971,proukakis2013quantum,Gladilin2019,Mei2019} to assess up to what point this PT can be described in terms of the standard BKT theory of equilibrium systems.
		From the early days of this field, dramatic consequences of non-equilibrium effects have been highlighted in polariton systems, from the non-trivial shape of the condensate in real and momentum spaces~\cite{Richard2005,Wouters2008} to the diffusive Goldstone mode in the collective excitation spectrum of polariton {condensates} with small polariton lifetime~\cite{szymanska2006nonequilibrium,carusotto2013quantum}. Furthermore, the possibility of breaking BKT algebraic decay of coherence in the quasi-ordered phase at very large distances has also been pointed out in~\cite{altman2015twodimensional}. Except for specific cases~\cite{dagvadorj2015nonequilibrium}, this occurs however on length scales well beyond the experimental possibilities.}
	{Still, it has been argued that for realistic system sizes the} {non-equilibrium character is responsible for an algebraic decay of the spatial coherence with an exponent exceeding the upper bound of 0.25 of equilibrium BKT theory} \cite{Roumpos6467,dagvadorj2015nonequilibrium} and {for} the ratio between spatial and temporal correlation exponents {being} equal to $2$~\cite{szymanska2006nonequilibrium,szymanska2007mean} {instead of $1$ as in the case of an equilibrium-like system~\cite{caputo2018}.}
	
	These theoretical predictions suggest that measurements of temporal coherence are a key ingredient to characterize the nature of the PT: while early works measured exponential or Gaussian decays of temporal coherence, {not compatible with a BKT transition~\cite{Roumpos6467,Krizhanovskii2006,Love2008,Kim2016,caputo2018,Askitopoulos2019}, and possibly related to single-mode physics \cite{AmelioCarusotto2020},} power-law decay of temporal correlations have been reported in recent works with improved samples~\cite{caputo2018}. Since the long polariton lifetime in Ref.~\cite{caputo2018} exceeds other characteristic time scales, one can reasonably assume the system to be in an equilibrium-like regime~\cite{caputo2018,ballarini2020directional}. On the other hand, to date there is no direct numerical or experimental measurement of a really non-equilibrium regime where the temporal and spatial algebraic exponents are different.
	
	Motivated by these open questions, in this work we undertake a detailed numerical study of the PT exhibited by an incoherently pumped (IP) $2d$ polariton {fluid} under realistic experimental parameters.
	We numerically investigate the non-equilibrium steady-state (NESS) phase {diagram} as a function of the pump power {and we characterize it} in terms  of the spatial and temporal correlations, {the spectrum of the collective excitation modes} and {the spatial distribution of} topological defects. {Our predictions shine new light on fundamental} properties of the {PT and on} its non-equilibrium nature.
	
	\section{Theoretical modelling}
	
	We describe the collective dynamics of the polariton fluid through a generalized stochastic Gross-Pitaevskii equation  {for the 2d polariton field as a function of the position $\bm{\mathrm{r}}=(x,y)$ and time $t$}, restricting our investigation here to the simplest case of a spatially homogeneous system with periodic boundary conditions. 
	The equation describes the  {effective} dynamics of the incoherently-pumped lower polariton field $\psi=\psi(\bm{\mathrm{r}},t)$~\cite{WC2007,carusotto2013quantum} {and includes the complex relaxation processes by means of a frequency-selective pumping source}~\cite{woutersLiew2010,chiocchetta2013,comaron2018dynamical}. The model, which can be derived from both truncated Wigner (TW) and Keldysh field theory~\cite{carusotto2013quantum,szymanska2006nonequilibrium} reads ($\hbar=1$):
	\begin{equation}
		\hspace{-4mm}id \psi = \bigg[ - \frac{
			\nabla^2}{2 m } +
		g|{\psi}|^2_{-} + \frac{i }{2}
		\bigg( \frac{P}{1+\frac{|{\psi}|^2_{-}}{n_\mathrm{s}}} -
		\gamma \bigg) \\  +\frac{1}{2}\frac{P}{\Omega}\frac{\partial}{\partial t} \bigg]
		\psi {\, dt}+  dW
		\label{eq:SGPE_pol}
	\end{equation}
	where $m$ is the polariton mass, {$g$ is} the polariton-polariton interaction strength, $\gamma$ {is the polariton loss rate (inverse of the polariton lifetime)}, {$P$ the strength of the incoherent pumping providing the gain, $n_\mathrm{s}$ is the saturation density, and $\Omega$ sets the characteristic scale of the frequency-dependence of gain}. The renormalized density $|{\psi}|^2_{-} \equiv \left(\left|{\psi} \right|^2 - {1}/{{(2dV)}} \right)$  {includes} the subtraction of the Wigner commutator contribution (where $dV=a^2$ is the {volume element} of {our $2d$ grid of} spacing $a$). 
	The zero-mean white Wiener noise $dW$ fulfils
	{$\left <dW(\textbf{r},t)dW(\textbf{r}^\prime,t)\right> = 0$, $\left < d W^*(\textbf{r},t)dW(\textbf{r}^\prime,t)\right> = [(P+\gamma)/2] \delta_{\textbf{r},\textbf{r}^\prime}dt, $}
	where the nonlinear density term is neglected since $|\psi|^2/n_s \ll 1$.
	To describe the physics of the model we start by considering Eq.~\eqref{eq:SGPE_pol} at a mean-field (MF) level, i.e. in the absence of the Wiener noise.
	As widely discussed in the literature~\cite{WC2007,WoutersCarusotto2010}, for the case of a frequency-independent pump ($\Omega = \infty$), a condensate with density $|\psi^\mathrm{SS}|^2 = n_\mathrm{s} \left({P}/{  \gamma} -1 \right)$ is expected to appear {for pump strengths above threshold $P > P_\mathrm{MF}= \gamma$} and to grow linearly in $P$ with a {slope determined by} the saturation density $n_\mathrm{s}$.
	For {a frequency-selective pump ($\Omega \neq \infty$), the NESS density loses its linear dependence on $P$, and takes the slightly more complicated form
		{$|\psi^{\mathrm{SS}}|^2 = n_\mathrm{s} \left[ {P}/ \left( P g|\psi^{\mathrm{SS}}|^2/ \Omega + \gamma \right)-1 \right]$~\cite{WoutersCarusotto2010}}.
		
		The effect of small excitations around the bare condensate steady-state solution can be described
		by means of {the linearized Bogoliubov} approximation~\cite{pitaevskii2003bose,carusotto2013quantum}.
		By linearizing the deterministic part of Eq.~\eqref{eq:SGPE_pol} around the steady state solution $\psi(\textbf{r},t) = \psi^\mathrm{SS} + \delta \psi(\textbf{r},t) e^{-i \omega t}$ we obtain a pair of coupled Bogoliubov equations for the field $\delta \psi(\textbf{r},t)$ and its complex conjugate $\delta\psi^*(\textbf{r},t)$. Thanks to translational invariance, the different $\textbf{k}$-modes are decoupled, so we can move to Fourier space and define a $\textbf{k}$-dependent Bogoliubov matrix $\mathcal{L}_\textbf{k}$~\cite{WoutersCarusotto2010,chiocchetta2013},
		\begin{equation}
			\mathcal{L}_\textbf{k} = \left(
			\begin{array}{cc}
				\Lambda ( {\epsilon_\textbf{k}} + \mu -i \Gamma )  & \Lambda (\mu -i \Gamma ) \\
				\Lambda^{*} (-\mu -i \Gamma  ) & \Lambda^{*} ( -{\epsilon_\textbf{k}}- \mu -i \Gamma ) \\
			\end{array}
			\right)
		\end{equation}
		with $\Gamma = {\gamma} \left( P-\gamma \right)/{2P}$, {the free-particle dispersion $\epsilon_\textbf{k} = k^2/2m$, the interaction energy $\mu = g |\psi^{\mathrm{SS}}|^2 $ and $\Lambda = \left( \gamma_a + i \gamma_b \right)$ with $\gamma_a={1}/{[1+\left({P}/{(2 \Omega)} \right)^2]}$ and $\gamma_b=-{P \gamma_a}/{2 \Omega} $.}
		The diagonalization of $\mathcal{L}_\textbf{k}$ eventually leads to the double-branched excitation spectrum
		\begin{equation}
			\begin{split}
				\omega_\textbf{{k}}^{\pm}  &= -i \left[  {\gamma_a \Gamma }-{\gamma_b} ({\epsilon_\textbf{{k}}} + \mu)  \right] \pm  \\
				&\sqrt{ \Gamma^2 {\gamma_a}^2 + {\gamma_b}^2 \mu^2 - 2 \Gamma \gamma_a {\gamma_b} \left( \epsilon_\textbf{{k}} + \mu \right) - {\gamma_a}^2  \epsilon_\textbf{{k}} \left( \epsilon_\textbf{{k}} + 2 \mu \right) }.  
				\label{eq:BogSpect}
			\end{split}
		\end{equation} 
		
		{At high momenta $\mathbf{k}$ this} spectrum {recovers} a single-particle behaviour with parabolic dispersion{, while the frequency-dependence of pumping results in an increasing linewidth for growing $\mathbf{k}$.}
		{For small $\textbf{k} \to 0$, the Goldstone mode describing long-wavelength twists of the condensate phase and associated to the spontaneously broken U(1) symmetry} exhibits {the} \textit{diffusive} behaviour typical of driven-dissipative systems \cite{carusotto2013quantum}, rather than {the sonic one}  characteristic of their equilibrium counterpart \cite{pitaevskii2003bose}.
		
		{This physics is illustrated in} Fig.~\ref{fig:Bogoliubov}, where {the prediction of} Eq.~\eqref{eq:BogSpect} is plotted {for increasing values of the pump strength $P$. The value of the critical momentum}
		\begin{equation}
			k_\mathrm{c} = \sqrt{ {2m} \left[  -   \frac{ \Gamma  {\gamma_b} }{{\gamma_a}}- \mu 
				+\frac{\sqrt{ \left(\Gamma ^2+\mu ^2\right) \left({\gamma_a}^2+{\gamma_b}^2\right)}}{{\gamma_a}}   \right] }
			\label{eq:kcrit}
		\end{equation}
		{separating} the diffusive behaviour from the sonic one at higher $\mathbf{k}$ increases as the system moves away from the threshold point $P_\mathrm{MF}$. 
		
		\begin{figure}
			\centering
			\onefigure[width=\linewidth]{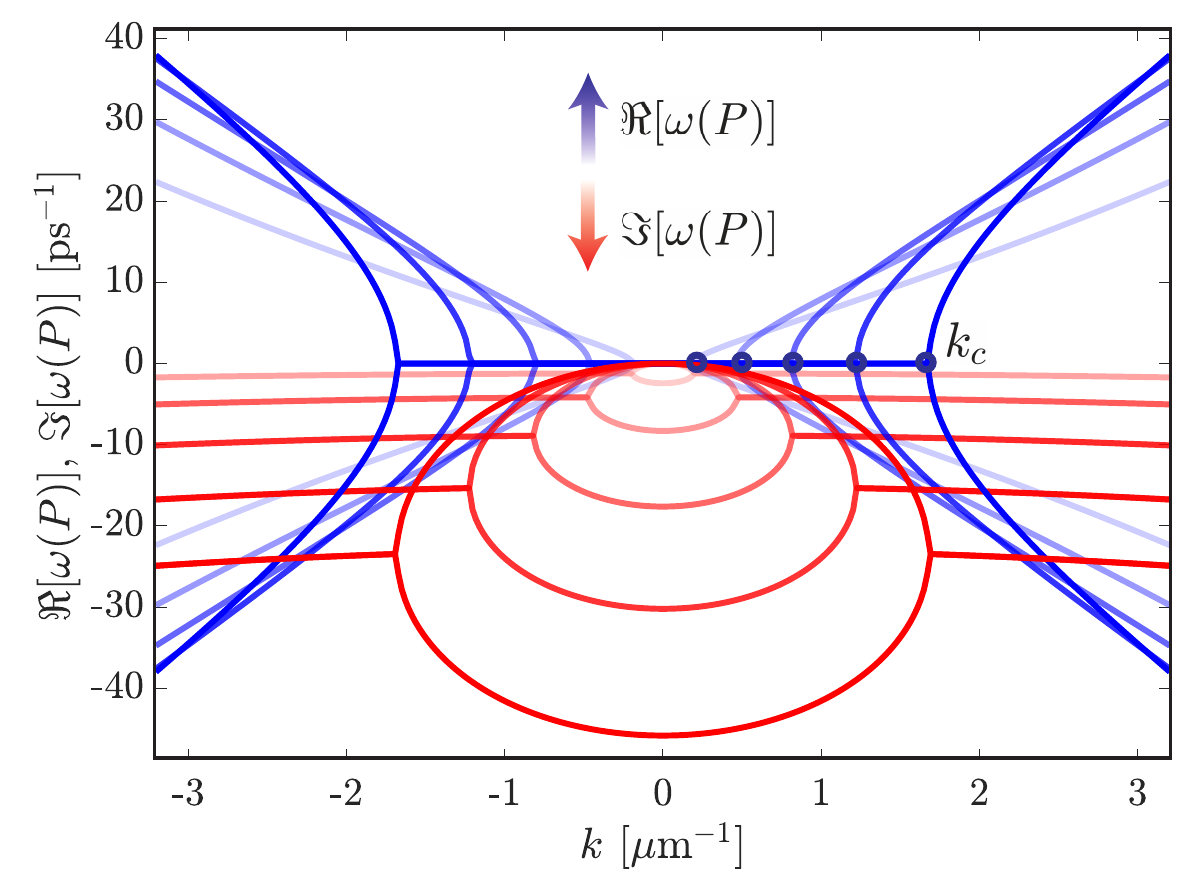}
			\caption{
				Real part (blue) and imaginary part (red) of the Bogoliubov excitation spectrum~\eqref{eq:BogSpect} calculated for the parameters {of the case IP$_{\Omega = 50}$ (whose density is plotted in Fig.~\ref{fig:phasediagram} as a blue curve), but 
					where the {relative} pump strength takes now the values $P/P_{\rm MF}~=~2,6,10,14,18,22$, increased as indicated by the colour gradients. 
					Corresponding real part for $P/P_{\rm MF}~=~1.06$ is shown in Fig.~\ref{fig:spectrum}(iv) as a blue curve.}
			}
			\label{fig:Bogoliubov}
		\end{figure}
		
		\section{The non-equilibrium Berezinskii-Kosterlitz-Thouless Phase diagram}
		\label{sec:PhaseD}
		
		We simulate the system dynamics by numerically integrating in time the stochastic differential equations for the polariton field {shown in}~\eqref{eq:SGPE_pol}; {numerical details are reported in \cite{SM}}.
		In Fig.~\ref{fig:phasediagram} we show {the typical} driven-dissipative BKT PT-diagram of an incoherently-pumped polariton condensate, in which {the different} observables are shown as a function of the {pump strength} $P$.
		{This is characterized by two} distinct phases: a) a disordered phase  displaying a low density of polaritons, an exponential decay of spatial correlations and a plasma of unbound free vortices; b) a superfluid phase  displaying a significant density of polaritons, an algebraic decay of spatial correlations and a low density of vortices, mostly bound in vortex-antivortex pairs~\cite{comaron2018dynamical,dagvadorj2015nonequilibrium,caputo2018}.  
		
		Our first {step} in the investigation of the IP polariton PT {was} to clarify the {impact} of fluctuations introduced by the stochastic noise {on the average density}. 
		The mean-field (stochastic) density $|\psi|^2=\int|\psi({r})|^2d{r}/L_xL_y$ [$ | \psi|^2 = |{\psi}|^2_{-}$] is calculated by evolving Eq.~\eqref{eq:SGPE_pol} without (with) the contribution of the {Wiener} noise. 
		{These two curves are} plotted in the inset of Fig.~\ref{fig:phasediagram} as dashed black and solid blue curves, respectively.
		Contrary to the mean-field case where $|\psi_\mathrm{MF}|=0$ in the disordered phase $P<P_\mathrm{MF}$, within the stochastic framework the density field is always non-zero, independently of the value of the pump $P$\footnote{In the disordered phase, fluctuations are responsible for building up a small but not negligible density {of incoherent polaritons}, the only zero-density point {coinciding} with a {vanishing pump strength $P = 0$}. {In} the quasi-ordered phase the {density} grows considerably {and asymptotically approaches} the mean-field prediction.}.
		
		{In contrast to the clear threshold shown by the MF curve, the smooth increase of the density with pump strength shown by the stochastic theory requires a more involved determination of the critical point.
			As done in previous works~\cite{dagvadorj2015nonequilibrium} --~and discussed in subsequent sections~-- our procedure to precisely determine the critical point involves the functional form of the} decay of correlation functions, {the} behaviour of vortices in the vicinity of the criticality and {the} appearance of the diffusive {Goldstone} mode in the spectrum.
		Interestingly, we note {in Fig.~\ref{fig:phasediagram}} that fluctuations are responsible for an {upward} shift {of the critical point $P_\mathrm{BKT}$ (vertical blue line) with respect to the MF value $P_{\mathrm{MF}}$  (vertical gray line).}
		{In order to unravel the dependence of $P_\mathrm{BKT}$ on the physical parameters $n_\mathrm{s}$ and $\Omega$, this figure} shows the phase diagram for three different choices of parameters, listed in the caption of the figure. 
		For each case analysed, the {critical point is} highlighted with a vertical coloured thick line.
		{As general trends, we find that} stronger fluctuations {in} higher modes ($\Omega\rightarrow\infty$) and smaller saturation {densities} ($n_\mathrm{s} \rightarrow 0$) lead to a larger shift of $P_\mathrm{BKT}$ with respect to the mean-field $P_\mathrm{MF}$.

		This feature can be understood by fixing one of the two parameters and focusing on the other.
		On the one hand, {for a frequency-independent} pump {($\Omega=\infty$, green and violet lines)}, we note that {increasing} $n_\mathrm{s}$ {makes} the BKT threshold $P_\mathrm{BKT}$ {to} shift closer to $P_\mathrm{MF}$: {the slope of the total density increases with $n_\mathrm{s}$, so} the critical density is reached at lower values {of the pump strength}.
		On the other hand, {for a fixed value of} $n_\mathrm{s} = 500 \mathrm{\mu m^{-2}}$ (blue and {violet} curves), {the presence of} a frequency-selective pump leads to {an effective thermal population of less field modes. As a consequence, a weaker pump is sufficient to concentrate a macroscopic population in the lowest modes, which has the effect of shifting the threshold point back towards the mean-field value $P_{\rm MF}$.}
		
		{As expected in a BKT-like picture, the IP phase transition can be pictorially understood as being mediated by the unbinding of vortex-anti-vortex pairs into a plasma of free vortices~\cite{dagvadorj2015nonequilibrium,altman2015twodimensional}.}
		In Fig.~\ref{fig:phasediagram} the NESS average number of topological defects $\left< N_\mathrm{v} \right>$ is plotted as a thick red line {for the parameters of the $\mathrm{IP_{\Omega=50}}$ case}. 
		{Details on the procedure we adopt to extract $N_\mathrm{v}$ are reported in \cite{SM} as well as the illustrations of three exemplary configurations of vortices across the BKT phase diagram}.
		The low-pump disordered phase is characterized by a {large} number of free vortices which are free to proliferate.
		As the pump power is increased and approaches the threshold point, the number of vortices $ \left< N_\mathrm{v} \right> \propto  \xi^{-1/2}$ {starts} to decay as expected for a continuum PT with diverging correlation length $\xi$, a detailed study of which is presented in Ref.~\cite{zamora2020}.
		At the critical point, around which the process of vortices pairing starts to take place, the average number of vortices is {still} non-zero {but these are mostly grouped in vortex-antivortex pairs}. 
		Due to vortex binding and annihilation processes, $\left< N_\mathrm{v}(P) \right>$ {shows a severe drop right above the critical point and rapidly decreases} to zero.
		Deep in the {quasi-condensed} phase when $P \gg P_\mathrm{BKT}$, as the stochastic density grows and onset of coherence appears, the dynamical annihilation processes are severe and eventually {leave} the system free of defects.
		
		\section{Spatial-temporal coherence and the critical region}
		\label{sec:decay_first_order}
		In this section we investigate the {long-distance, late-time} decay of the spatial and temporal first-order correlation functions, $g^{(1)}(\Delta r)$ and $g^{(1)}(\Delta t)$ respectively.
		We focus here on the results {for the IP$_{\Omega = 50}$ case}; the complementary study for the {IP$_{\Omega = \infty}^{n_\mathrm{s}=1500}$ case with a frequency-independent pump is illustrated in~\cite{SM}}.
		Within our semi-classical model, the spatial and temporal two-point first order correlation functions {are defined}, respectively, as
		\begin{eqnarray}
			g^{(1)} (\Delta r)  =\frac{\langle \psi^* ({r_0} + \Delta r,t) \psi^{} ({r_0},t)
				\rangle 
			}{\sqrt{ \langle \left| \psi({r_0} + \Delta  {r},t)\right|^2
				\rangle \langle \left| \psi({r_0},t) \right|^2
				\rangle}}\; ,
		\label{eq:corre_space} \\
		g^{(1)} (\Delta t) 	=
		\frac{\langle \psi^* ({r_c},t_0) \psi^{} ({r_c},t_0 + \Delta  t) \rangle }{\sqrt{\langle \left| \psi({r_c},t_0)\right|^2	\rangle \langle \left| \psi({r_c},t_0+ \Delta  t) \right|^2 \rangle}}\; ,
		\label{eq:corre_time}
	\end{eqnarray}
	and are calculated at a {sufficiently late} time $t = t_\mathrm{SS}$ {at which the system has reached} its NESS, and with $r_c = (L_x/2,L_y/2)$ {being} the central point of the {spatial grid}.
	The {numerical results for the spatial and temporal correlations} are illustrated in {panels a) and b) of Fig.~\ref{fig:correlators}, respectively.}
	
	Inspired by earlier works~\cite{dagvadorj2015nonequilibrium,Comaron2019}, we {characterize}  the behaviour of the steady-state correlation functions as function of the {pump strength $P$}.
	In Fig.~\ref{fig:correlators} we show the transition from an exponential decay $g^{(1)} \sim e^{-r/\xi}$ in the disordered phase, to a power-law decay $g^{(1)} \sim r^{-\alpha}$ in the quasi-ordered phase, as expected for the spatial correlation function of an equilibrium BKT transition. {The same behaviour is found for the temporal correlation function\footnote{Note that in our simulations the accessible time duration are not long enough to observe the finite-size-induced Schawlow-Townes decay~\cite{keeling2010keldysh,AmelioCarusotto2020}.}}.
	In order to identify whether {a given correlation function} is characterised by either exponential or algebraic decay, we {have fitted} each curve with both functions, paying particular attention to ensure that all computational results are correctly converged within the spatial and temporal windows selected for the fitting procedure~\cite{SM}.
	{We have then calculated the Root-mean-square} deviation (RMSD) of the residuals of the fits within the fitting window selected {and we have selected the fit that minimizes the RMSD \cite{SM}. In Fig.~\ref{fig:correlators} we superimpose on top of each correlation function $g^{(1)}$, the {most accurate} fitting curve, represented by red or blue dashed lines in the exponential or power-law cases, respectively.}
	
	\begin{figure}
		\centering
		\onefigure[width=\linewidth]{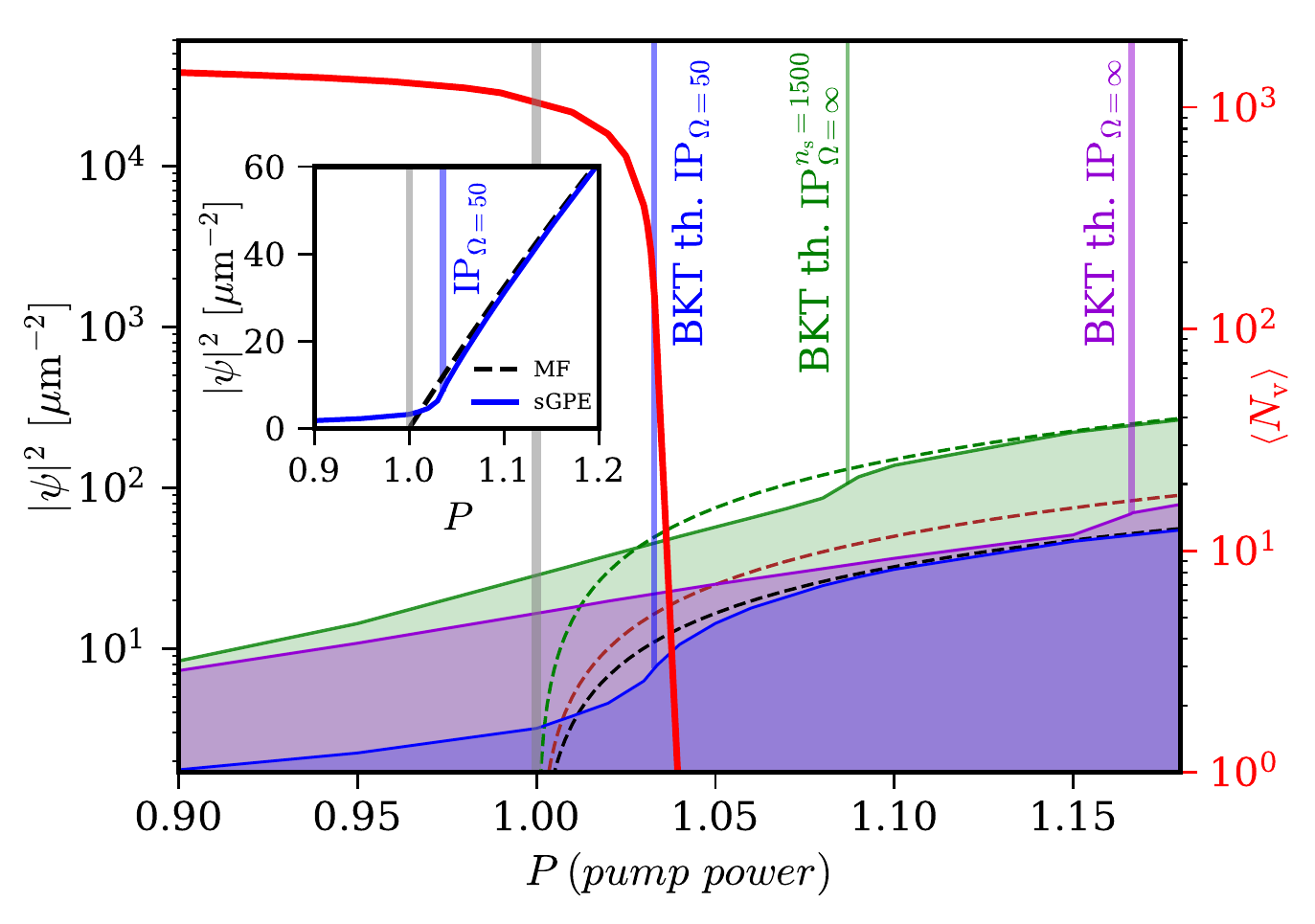}
		\caption{
			Non-equilibrium steady-state phase diagram showing mean-field (MF) and averaged stochastic density (sGPE) (dashed and solid coloured curves, respectively) in logarithmic-linear scale. 
			For each set of parameters, we associate a colour. {Blue} (labelled as IP$_{\Omega = 50}$)$: \Omega = 50 \gamma = 11.09 ps^{-1}$ and $n_\mathrm{s} = 500 \mathrm{\mu m^{-2}}$. {Green} (labelled as IP$_{\Omega = \infty}^{n_\mathrm{s}=1500}$): 
			$\Omega = \infty$ and $n_\mathrm{s} = 1500 \mathrm{\mu m^{-2}}$.  {Violet} (labelled as IP$_{\Omega = \infty}$): $\Omega = \infty$ and $n_\mathrm{s} = 500 \mathrm{\mu m^{-2}}$.
			{For each set of parameters, the BKT threshold} is shown as a vertical coloured line. {The vertical} {grey} line {shows} the mean-field threshold $P_\mathrm{MF} = 1$.
			The average number of vortices $\left< N_\mathrm{v}\right>$ for the IP$_{\Omega = 50}$ case {is} depicted as a red curve.
			The inset shows comparison between {the} mean-field (dashed black {line}) and {the} averaged stochastic density (blue solid {line}) for the IP$_{(\Omega = 50)}$ case, plotted in linear-linear scale.
		}
		\label{fig:phasediagram}
	\end{figure}

	\begin{figure}
		\centering
		\includegraphics[width=\linewidth]{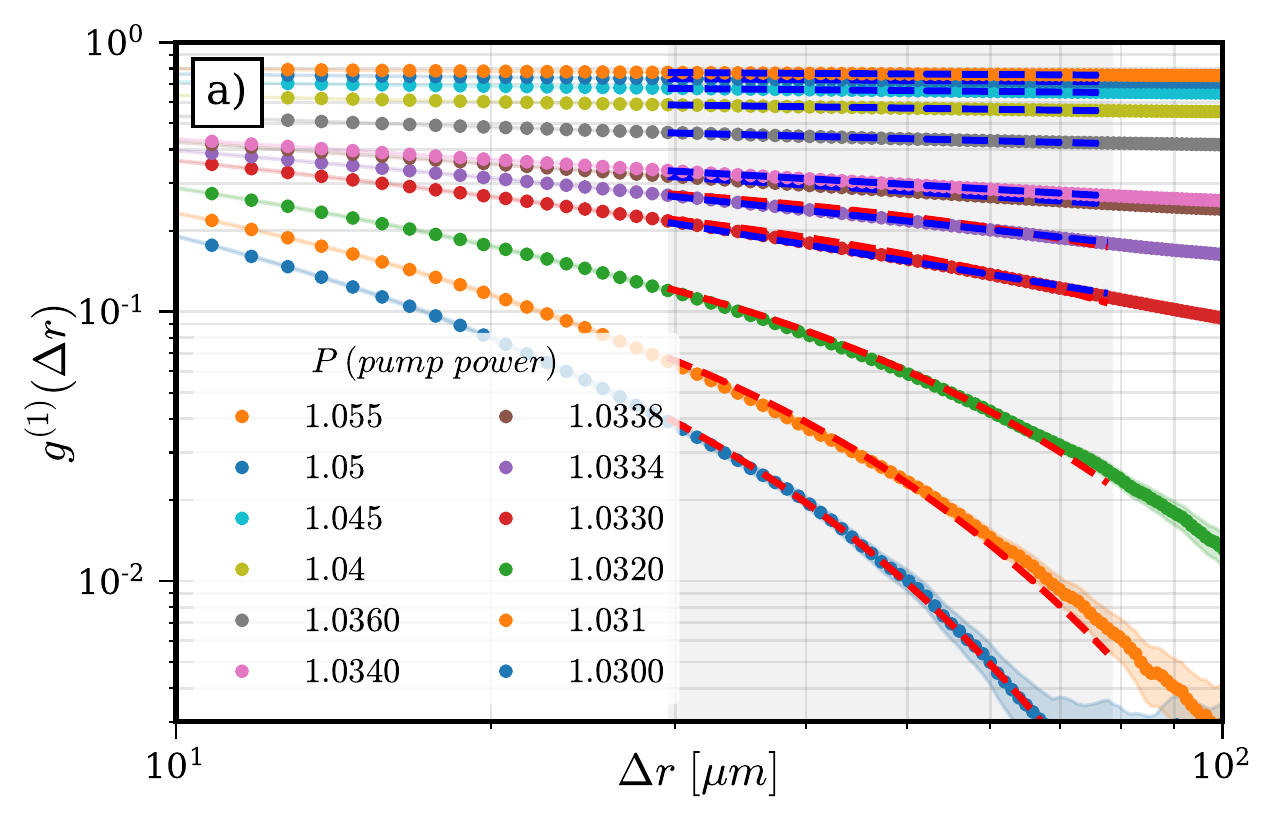}
		\includegraphics[width=\linewidth]{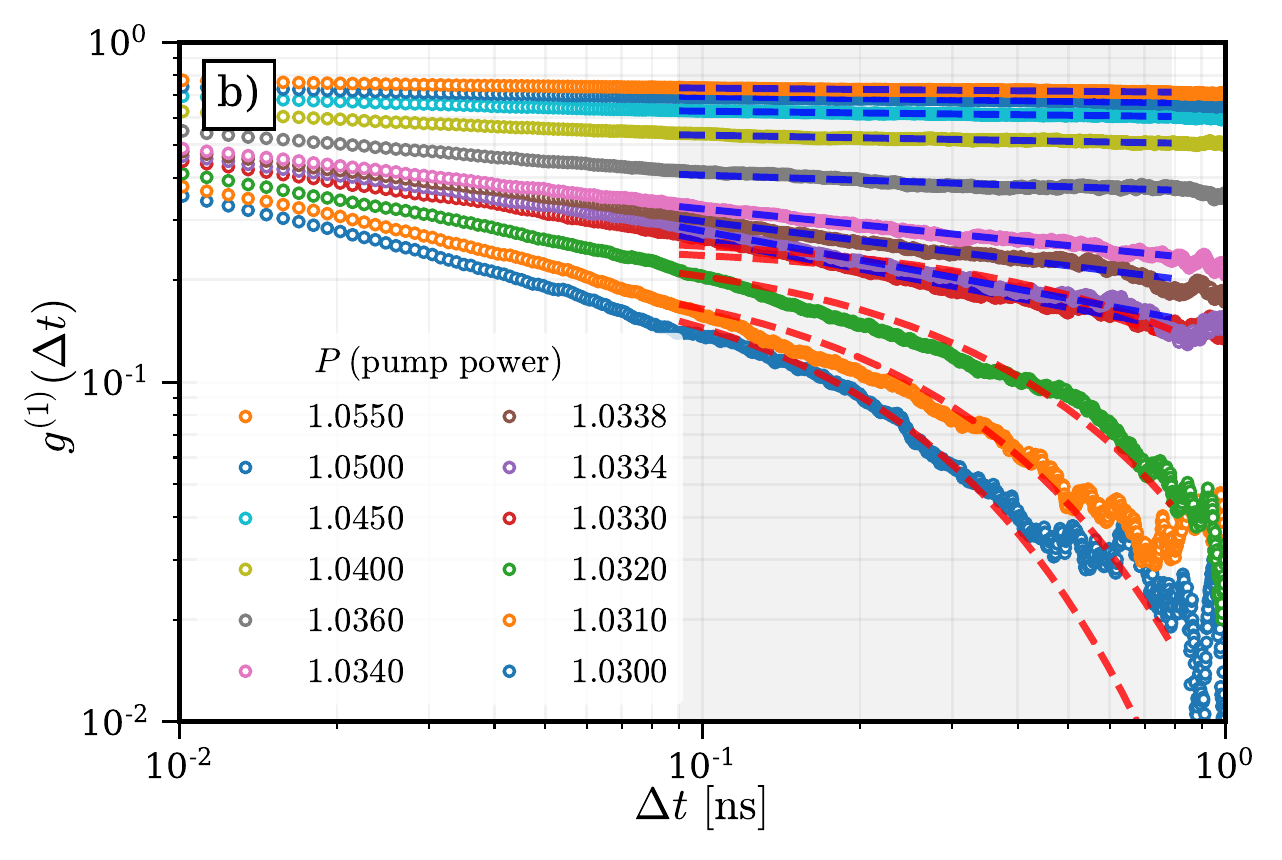}
		\caption{
			{Crossover from exponential to algebraic decay of the spatial} \textbf{(a)} and temporal \textbf{(b)} correlation functions, defined as in Eqs.~\eqref{eq:corre_space} and \eqref{eq:corre_time}.
			Thick dashed red (blue) curves correspond to exponential (power-law) fitting, from which {the} values of {the} correlation length $\xi$ and {of the} power-law exponents $\alpha_\mathrm{s}$ and $\alpha_\mathrm{t}$ plotted in Fig.~\ref{fig:exponents} {were extracted}.
			For each curve, we superimpose only the {best fitting option. Both fits are only shown for the curves which lie in the critical region. The fits are restricted to the chosen fitting window, indicated by the gray shadow.}
		}
		\label{fig:correlators}
	\end{figure}

	\begin{figure}[t!]
		\centering
		\includegraphics[width=\linewidth]{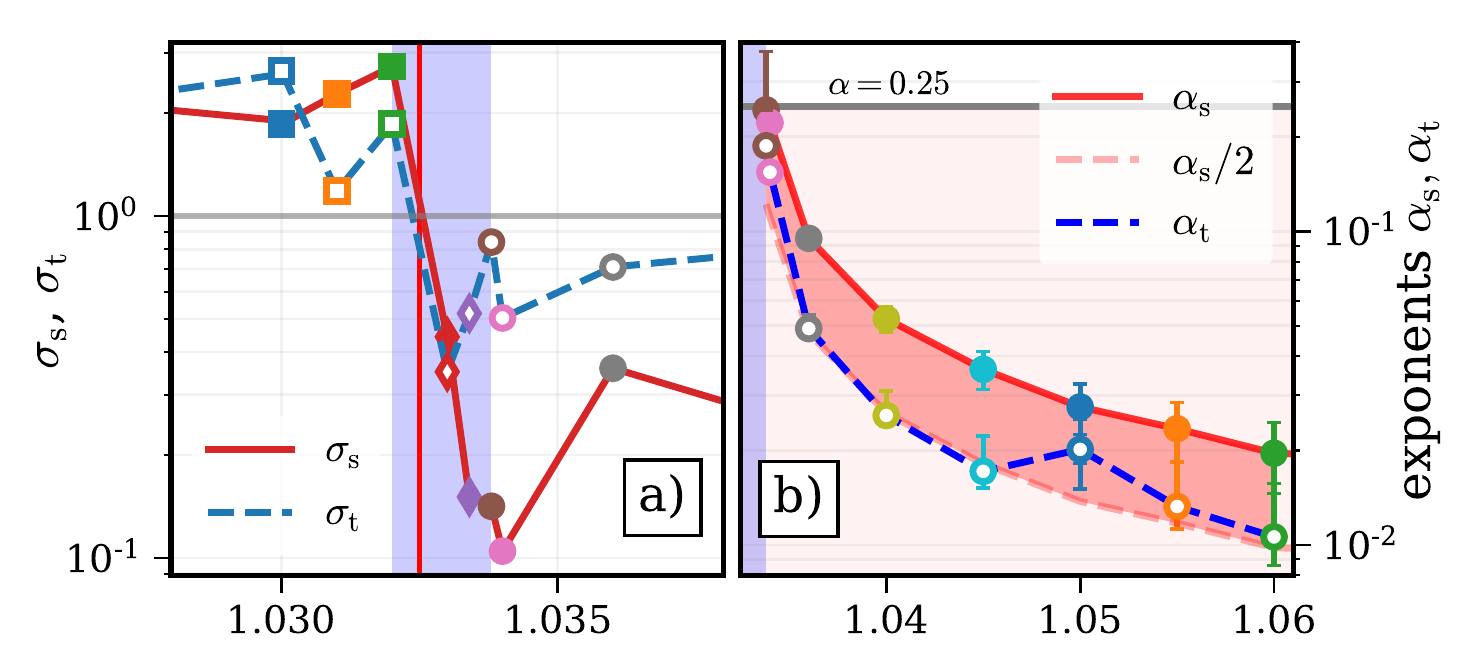}
		\includegraphics[width=\linewidth]{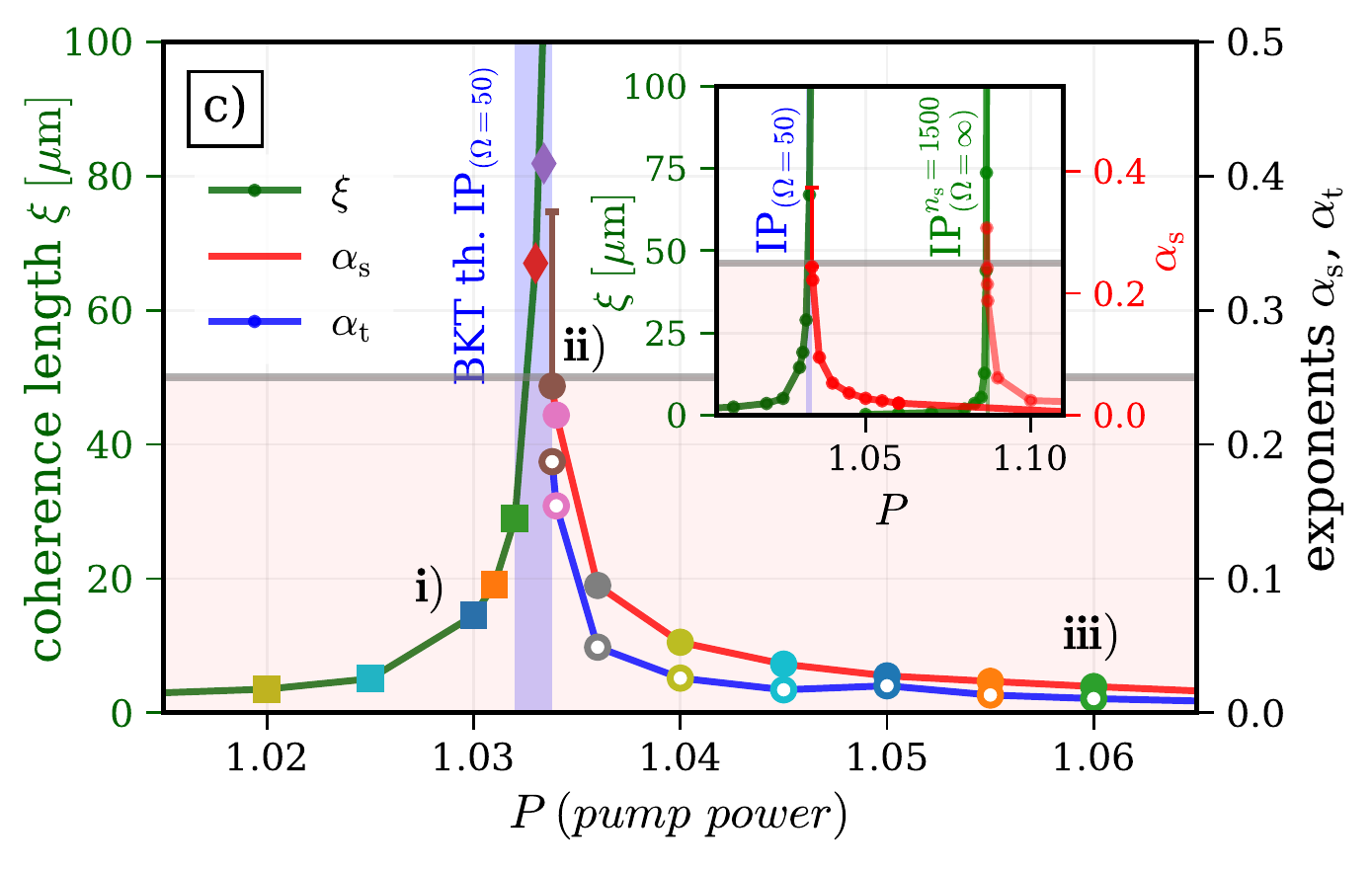}
		\caption{
			\textbf{a)}
			{For the IP$_{\Omega = 50}$ case, plot of the quantities $\sigma_\mathrm{s}$  (solid red curve) and $\sigma_\mathrm{t}$ (dashed blue line) defined  in Eqs.~\eqref{eq:RMSD}, which identify} the critical region (shaded blue region) and the critical point $P_\mathrm{BKT}$ (vertical red line).
			Squares, diamonds and circles correspond to points which fall before, within and above the critical region, respectively. 
			\textbf{b)}
			{Log-linear plot of the spatial} ($\alpha_\mathrm{s}$, {filled circles) and temporal ($\alpha_\mathrm{t}$, empty circles) exponents, extracted from the power-laws fits of Fig.~\ref{fig:correlators} with error bars.}
			\textbf{c)} The correlation length $\xi$ (green solid curve) diverges when approaching $P_\mathrm{BKT}$ {from} the disordered phase.
			{Away from the critical point into} the quasi-ordered phase, {the decay of the} spatial (temporal) algebraic exponent $\alpha_\mathrm{s}$ ($\alpha_\mathrm{t}$) is shown as empty (filled) circles and red (blue) solid line.
			{The excitations spectrum for three characteristic values of the pump strength indicated with} \textbf{i)}, \textbf{ii)} and \textbf{iii)} is shown in Fig.~\ref{fig:spectrum}.
			{The inset shows a plot of} $\xi$ and $\alpha_\mathrm{s}$ {for the} IP$_{\Omega = 50}$ and IP$_{\Omega = \infty}^{n_\mathrm{s}=1500}$ cases.
			In all panels above, the {colour of the markers corresponds to the one  of the different curves in Fig.~\ref{fig:correlators}.}
		}
		\label{fig:exponents}
	\end{figure} 
	\begin{figure*}[]
		\centering
		\includegraphics[width=.99\linewidth]{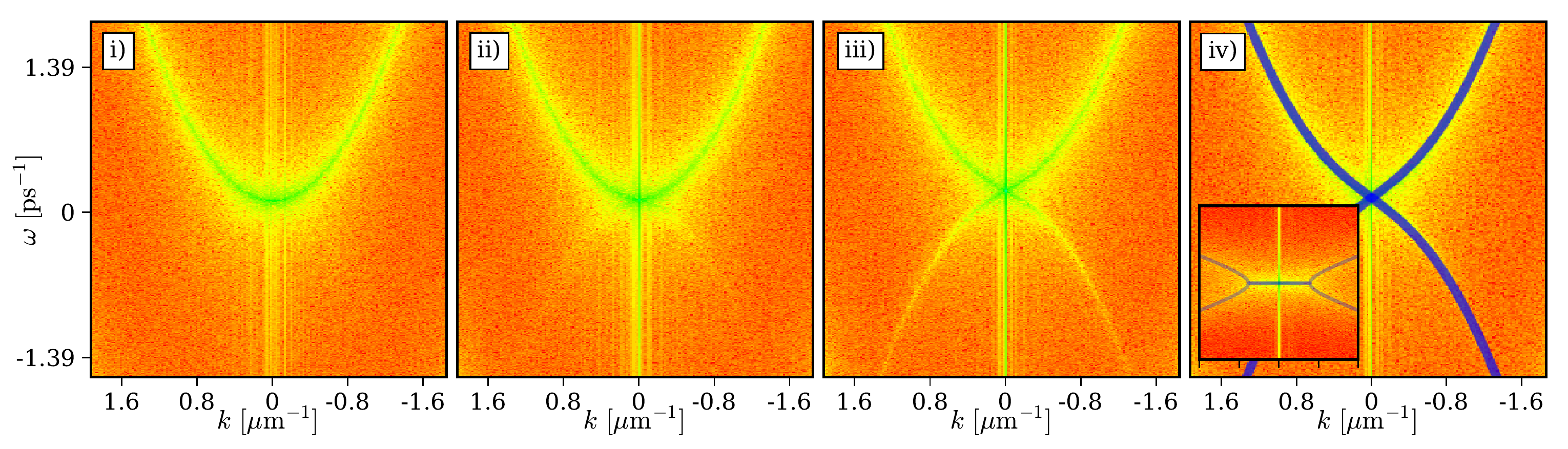}
		\caption{
			Panels i-iii) show colorplots of the spectra obtained numerically from the Fourier transform of $|\psi{(\textbf{r}, t)}|^2$, for {the three points} {i)} $P=1.03$, {ii)} $P=1.0338$ and {iii)} $P=1.06$ highlighted in Fig.~\ref{fig:exponents}.
			Panel iv) shows a comparison with {the real part of} the analytical dispersion in \eqref{eq:BogSpect} (blue curves) for case {iii)}.
			{The inset shows a different numerical simulation with a ($100\times$) larger box and spatial discretization, able to capture the  low-$k$ region and the diffusive branch of the spectrum.}
		}
		\label{fig:spectrum}
	\end{figure*} 
	
	Fig.~\ref{fig:exponents}(a) shows how we characterise the critical region by means of the RMSD ratios of the fits {of the spatial (red solid curve) and temporal (dashed blue curve) correlators,} namely
	\begin{equation}
		\sigma_\mathrm{s} = \frac{\mathrm{RMSD}^\mathrm{pow}_\mathrm{s}}{\mathrm{RMSD}^\mathrm{exp}_\mathrm{s}}, \qquad \sigma_\mathrm{t} = \frac{\mathrm{RMSD}^\mathrm{pow}_\mathrm{t}}{\mathrm{RMSD}^\mathrm{exp}_\mathrm{t}}\,.
		\label{eq:RMSD}
	\end{equation}
	{By visually} comparing the residuals of the exponential and power-law fits {on this figure}, one {can} infer {the position of the critical point} as 
	{the point where the two curves go through $1$. This point indicates the exponential-to-power-law transition, which takes place for the same $P_\mathrm{BKT}~\sim~1.0325$ (vertical red solid line in Fig.~\ref{fig:exponents}) for both spatial and temporal correlation functions}\footnote{ 
		In both Fig.~\ref{fig:correlators} and Fig.~\ref{fig:exponents}(a), there exists an intermediate regime in an interval of  $P_\mathrm{BKT}$, where the curves are neither exactly fitted by a power-law or an exponential form. 
		Therefore, we refer to the critical region as the portion of the phase diagram {located in between} the last correlator showing ``clear'' exponential decay (lower bound) and the first exhibiting ``clear'' power-law decay (upper bound). {In our case, these lower and upper bounds 
			are located at} $P = 1.032$ and $P = 1.0338$, respectively. In both Figs.~\ref{fig:phasediagram}~and~\ref{fig:exponents}, the critical region is highlighted with a blue shading.}.
	
	{This analysis of the decay of the correlation functions} allows us to extract quantities which are strictly linked to the physical nature of the PT. Namely, the correlation length $\xi$, extracted from the exponential fit {in the disordered phase}, and the algebraic exponents $\alpha_s$, $\alpha_t$ which quantify the algebraic decay of space and time correlators {in the quasi-ordered one. In equilibrium systems, the former} is known to be related to the superfluid density {\cite{Berezinskii1973}}.
	These quantities are plotted in Fig.~\ref{fig:exponents}(c), as a function of the {pump strength} $P$ and represented as solid green [$\xi(P)$], solid red [$\alpha_\mathrm{s}(P)$] and blue thick [$\alpha_\mathrm{t}(P)$] curves. 
	Markers in Fig.~\ref{fig:exponents} are coloured {in a way} to match {the ones} of Figs. \ref{fig:correlators}. 
	In the disordered phase [left part of Fig.~\ref{fig:exponents}(c)] the coherence length $\xi(P)$ diverges when approaching the critical region from the left, as expected for a continuum PT (in finite systems) undergoing critical slowing down\footnote{While the dataset extracted is suitable for a qualitative description of the PT, a possible quantitative extraction of critical exponents would require a more advanced scaling analysis with larger sample sizes.}.
	In the quasi-ordered phase [right part of Fig.~\ref{fig:exponents}(b-c)], the exponents $\alpha_\mathrm{s}$ and $\alpha_\mathrm{t}$ show a decreasing behaviour as the control parameter $P$ is increased, {which is connected} to the expected onset of coherence.
	
	In Fig.~\ref{fig:spectrum} we plot three exemplary cases of excitation spectrum calculated from the {spatio-temporal} Fourier Transform (FT) of $|\psi{(\textbf{r}, t)}|^2$ across the PT. 
	As expected, the system moves from a free-particle quadratic dispersion below the transition [Fig.~\ref{fig:spectrum}(i)] to a non-equilibrium spectrum, as in \eqref{eq:BogSpect}, above the transition [Fig.~\ref{fig:spectrum}(iii-iv)].
	We find the analytical Bogoliubov dispersion~\eqref{eq:BogSpect} to correctly describe our numerics: {the agreement between the peak of the numerical spectrum (colour map) and the analytical prediction (blue curves) is explicitly illustrated for the {last} case in Fig.~\ref{fig:spectrum}(iv).}
	{{For this case, we find that {the} critical momentum $k_c(P=1.06) = 1.26 \times 10^{-2} \mathrm{\mu m^{-1}}$ is 
			on the order of the momentum discretization $\Delta k = \pi/L=1.06 \times 10^{-2} \mathrm{\mu m^{-1}}$ of the numerical simulation; as a consequence, in the main panels the diffusive branch is hidden by the sonic behaviour of the dispersion at $k>k_c$.
			The numerical value $k_c^\textrm{num} = 1.0(5) \times 10^{-2} \mathrm{\mu m^{-1}}$ is extracted by simulating a system with a ($100 \times$) smaller $\Delta k$ \cite{SM}: the low-$k$ part of the spectrum, plotted as an inset of Fig.~\ref{fig:spectrum}(iv), is now visible and in good agreement with the analytically predicted curve.}}
	
	\section{Discussion on the nature of the phase transition}
	\label{sec:detPbkt}
	
	Previous experimental \cite{Roumpos6467} and theoretical \cite{dagvadorj2015nonequilibrium} works showed that a spatial power-law exponent exceeding the $\alpha_\mathrm{s} =0.25$ {upper bound of the equilibrium theory is a signature of the non-equilibrium nature of the PT.
		While this is evidently the case for the IP$_{\Omega=\infty}^{n_\mathrm{s}=1500}$ simulations, 
		the numerically obtained value of the exponent in the IP$_{\Omega=50}$ case never exceeds the equilibrium upper bound.}
	However, by enlarging the system size we find that $g^{(1)}({\Delta}r)$ is converged in space over all the quasi-ordered pump range {except for the extreme point $P = 1.0338$.} This is expected, as {in} the very vicinity of criticality { finite size effects can be most} important.
	As shown in \cite{SM}, by enlarging the box by 1.5 and 2 times, power-low exponents are found to lie within the interval $0.25 < \alpha_s < 0.35$ [reported in Fig.~\ref{fig:exponents}(b,c) as a large errorbar for the brown point $\alpha_\mathrm{s}(P=1.0338)$]. {This confirms} the argument that for non-equilibrium driven-dissipative system $\alpha_\mathrm{s}$ {can exceed} the upper equilibrium limit of $\alpha=0.25$ {in the critical region, for both frequency-independent and frequency-dependent pumping.}
	
	{A key difference between equilibrium and non-equilibrium PTs is encoded in the relation between the $\alpha_\mathrm{s}$ and $\alpha_\mathrm{t}$ exponents. In the equilibrium case, the sonic nature of the  dispersion leads to $\alpha_\mathrm{s} = \alpha_\mathrm{t}$. For a non-equilibrium driven-dissipative condensate, the diffusive nature of the Goldstone mode suggests instead that $\alpha_\mathrm{s} \sim 2 \alpha_\mathrm{t}$ \cite{szymanska2006nonequilibrium}.} 
	{At first sight, the prominent sonic branch visible in the spectrum of Fig.~\ref{fig:spectrum}(iii) could suggest that we are in a similar equilibrium-like scenario as in Ref.~\cite{ballarini2020directional}, where almost equal values were measured for $\alpha_\mathrm{s}$ and $\alpha_\mathrm{t}$, in strong contrast to our numerics. Looking at the excitation spectrum in Refs.~\cite{caputo2018,ballarini2020directional} reveals that the critical momentum $k_c(P=1.06)$ is there $2.53 \times 10^{2}$ times smaller than the one considered here, giving a characteristic length $2\pi/k_c$ that largely exceeds the system size. This is due to the much longer lifetime displayed by polaritons in those experiments and is responsible for the absence of an observable diffusive region in the Goldstone mode.}
	{Our numerical study shows instead a power-law decay of both spatial and temporal correlation function, with an exponent ratio $\alpha_\mathrm{s} \sim 2 \alpha_\mathrm{t}$ [Fig.~\ref{fig:exponents}(b)], suggesting a non-equilibrium nature of the condensate.
		However, due to the inability to numerically simulate a large enough box to clearly highlight the diffusive Goldstone mode, we cannot determine whether the different values measured for $\alpha_\mathrm{s}$ and $\alpha_\mathrm{t}$ is due to its non-equilibrium nature or finite-size effects, or an interplay between the two.}

	\section{Conclusions}
	
	In this paper we {have undertaken} a detailed numerical analysis to {investigate} the {non-equilibrium} phase transition {displayed by} a polariton {system} under incoherent pumping. 
	We {have characterized} the non-equilibrium phase diagram within both mean-field and stochastic pictures, confirming {for realistic system sizes} a BKT-like scenario for non-equilibrium condensates featuring a crossover between binding/unbinding of vortices and between an exponential/power-law decay of correlations.
	Particular attention was given to the role of fluctuations in the shift of the critical point with respect to the mean-field picture and to the {long-distance and late-time decay of the spatial and temporal correlation functions}.
	Our findings show that the non-equilibrium driven-dissipative phase transition exhibits an algebraic exponent exceeding the upper-bound equilibrium limit of $1/4$ in agreement with previous experimental \cite{Roumpos6467} and theoretical \cite{dagvadorj2015nonequilibrium} works.
	A non-equilibrium nature of the condensate is also suggested by the ratio $\alpha_\mathrm{s} /  \alpha_\mathrm{t}\sim 2$ of the algebraic decay exponents of space and time correlators, extracted by our numerical simulations and suggested by analytical {calculations within a Keldysh} framework~\cite{szymanska2006nonequilibrium}. 
	We note however that such an effect could be also due to a possible interplay with finite-size effects.
	It would be of a great interest to explore the interplay between non-equilibrium and finite-size effects in spatial correlations in future works.
	Our results suggest that {a complete} characterization of the {non-equilibrium Berezinskii-Kosterlitz-Thouless} phase transition is within current experimental reach {using polariton fluids}.
	
	\acknowledgments
	We acknowledge financial support from the Quantera ERA-NET cofund projects NAQUAS (through the Engineering and Physical Science Research Council (EPSRC), Grant No. EP/R043434/1 (P.C. and N.P.P.)), and InterPol (through the National Science Center, Poland, Grant No. 366 2017/25/Z/ST3/03032 (P.C.) and the EPSRC  Grant No. 368 EP/R04399X/1 (M.H.S.)). We acknowledge EPSRC Grant No. EP/K003623/2 (M.H.S.) and support from the Provincia Autonoma di Trento and from the Quantum Flagship Grant PhoQuS (820392) of the European Union (I.C.).

\pagebreak

\cleardoublepage

\begin{center}
	\textbf{\large Supplementary Material for: Non-equilibrium Berezinskii-Kosterlitz-Thouless transition in driven-dissipative condensates}
\end{center}

\vspace{8mm}
%

	In this Supplementary Material we present technical details related to our numerical procedure and analysis for the study of a non-equilibrium driven-dissipative phase transition. We discuss important numerical details, numerical methods and convergence of spatial and temporal correlations and we explain numerical methods for topological defect counting. Finally, show a detailed analysis on the long-range behaviour close to the critical point.

\

\section{Numerical calculations}
We simulate the system dynamics by numerically integrating in time the stochastic differential equations for the polariton field {shown in} Eq.~\blue{(1)} of the main text.
The numerical integration is performed on a 2d-grid with periodic boundary conditions. 
The lattice is composed of $301^2$ grid points, with total lengths $L_\mathrm{x} =L_\mathrm{y}= 295.11 \mu m$ and lattice spacing $a=0.98\mu m$. 
Notice that the lattice spacing $a$, {both} introduces a cut-off $\propto a^{-1}$ in the momentum representation of the field, and  {is chosen} between a lower bound given by the macroscopic scale of the system, such as the healing length, and the upper bound $a^2 \gg g/\gamma$ given by the validity of the truncated Wigner methods used for the description of the stochastic field equations~\cite{Sinatra2008,carusotto2013quantum,comaron2018dynamical}.
The time dynamics of the polariton field is here performed by integrating~Eq.~\blue{(1)} in time with the XMDS2 software framework~\cite{dennis2013xmds2}. Here we have used a fourth-order Runge-Kutta algorithm with fixed time-step of $4.5\cdot 10^{-2} \mathrm{ps}$, which ensures stochastic noise consistency.
All the results expressed in the present work are converged with respect to the number of stochastic realisations $\mathcal{N} = 200$.
%

{
	In the present study, we use typical experimental parameters {\cite{nitsche2014algebraic}}: lifetime $\tau = 1/\gamma= 4.5 \mathrm{ps}$, $m = 6.2 \ 10^{-5} \ m_\mathrm{e}$, $g = 6.82 \ 10^{-3} \ \mathrm{meV \mu m^2} $. 
	In Fig.~1 of the main text the non-equilibrium steady-state properties of the incoherently pumped polariton system are discussed in term of three different sets of the values ($\Omega $, $n_\mathrm{s}$). Such sets are labelled as
	IP$_{\Omega = 50}$ ($\Omega = 50 \gamma = 11.09 ps^{-1}$ and $n_\mathrm{s} = 500 \mathrm{\mu m^{-2}}$);
	IP$_{\Omega = \infty}^{n_\mathrm{s}=1500}$ ($\Omega = \infty$ and $n_\mathrm{s} = 1500 \mathrm{\mu m^{-2}}$) and  
	IP$_{\Omega = \infty}$ ($\Omega = \infty$ and $n_\mathrm{s} = 500 \mathrm{\mu m^{-2}}$).
	The analysis on the coherence properties in the main text is mainly centred around the case IP$_{\Omega = 50}$ while the complementary analysis for the case IP$_{\Omega = \infty}^{n_\mathrm{s}=1500}$ is presented in this supplemental document. 
	For the sake of clarity, it is also worth noting that these sets of parameters have been employed in other works of the authors.
	Specifically, in Ref.~Comaron~\textit{et. al}~\cite{comaron2018dynamical} the set of parameters IP$_{\Omega = 50}$ and IP$_{\Omega = \infty}^{n_\mathrm{s}=1500}$ have been used for the investigation of late-time quench dynamics properties of driven-dissipative condensate, while in Zamora~\textit{et. al}~\cite{zamora2020}, the validity of the Kibble-Zurek hypothesis for microcavity fluids is discussed; in the latter only the case IP$_{\Omega = 50}$ is considered.
}

The steady state values of the observables discussed in the analysis of the main paper have been reached after evolving the MF (sGPE) model from a very small initial value of the density (from a Gaussian white noise) to a steady state with a non-zero coherent density, thereby simulating all cases starting from a state without a condensate.


\section{First-order spatial and temporal correlations: methods and numerical convergence}
\label{app:appendix_convergence}

In Fig.~\ref{fig:correlations_all_scales} we repropose the data shown in Fig.~\blue{3} of the main text, plotted both in log-log (left) and log-linear (right) scale.
\begin{figure*}[t!]
	\centering
	\includegraphics[width=.455\linewidth]{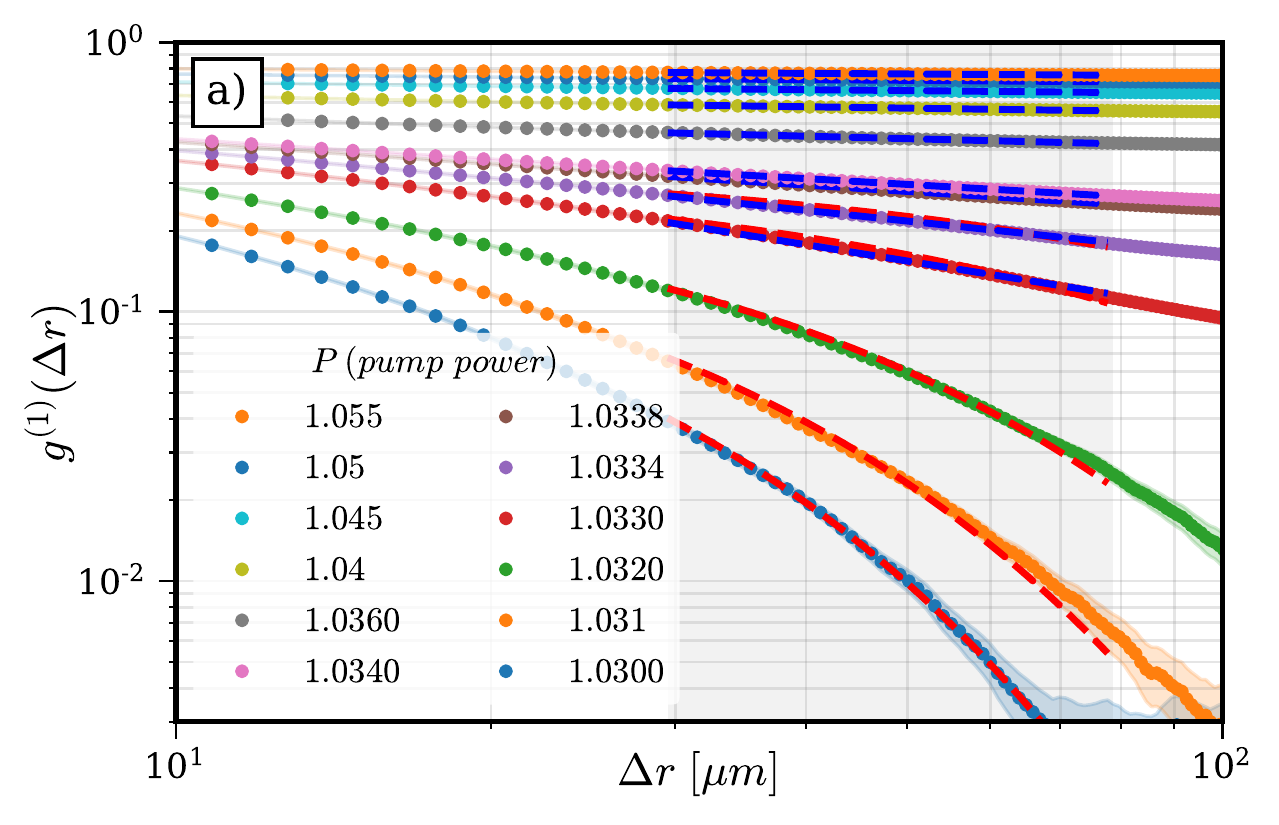}
	\includegraphics[width=.445\linewidth]{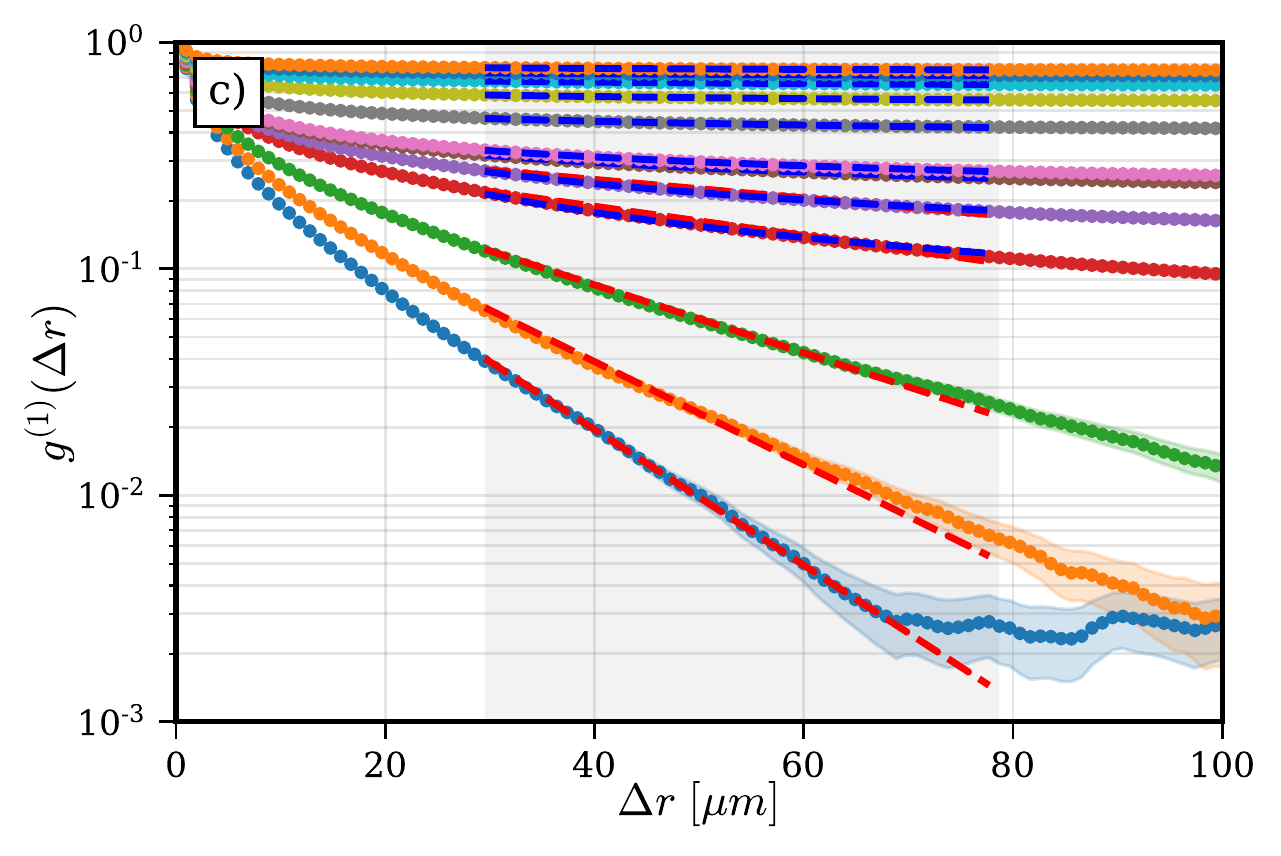}
	\includegraphics[width=.45\linewidth]{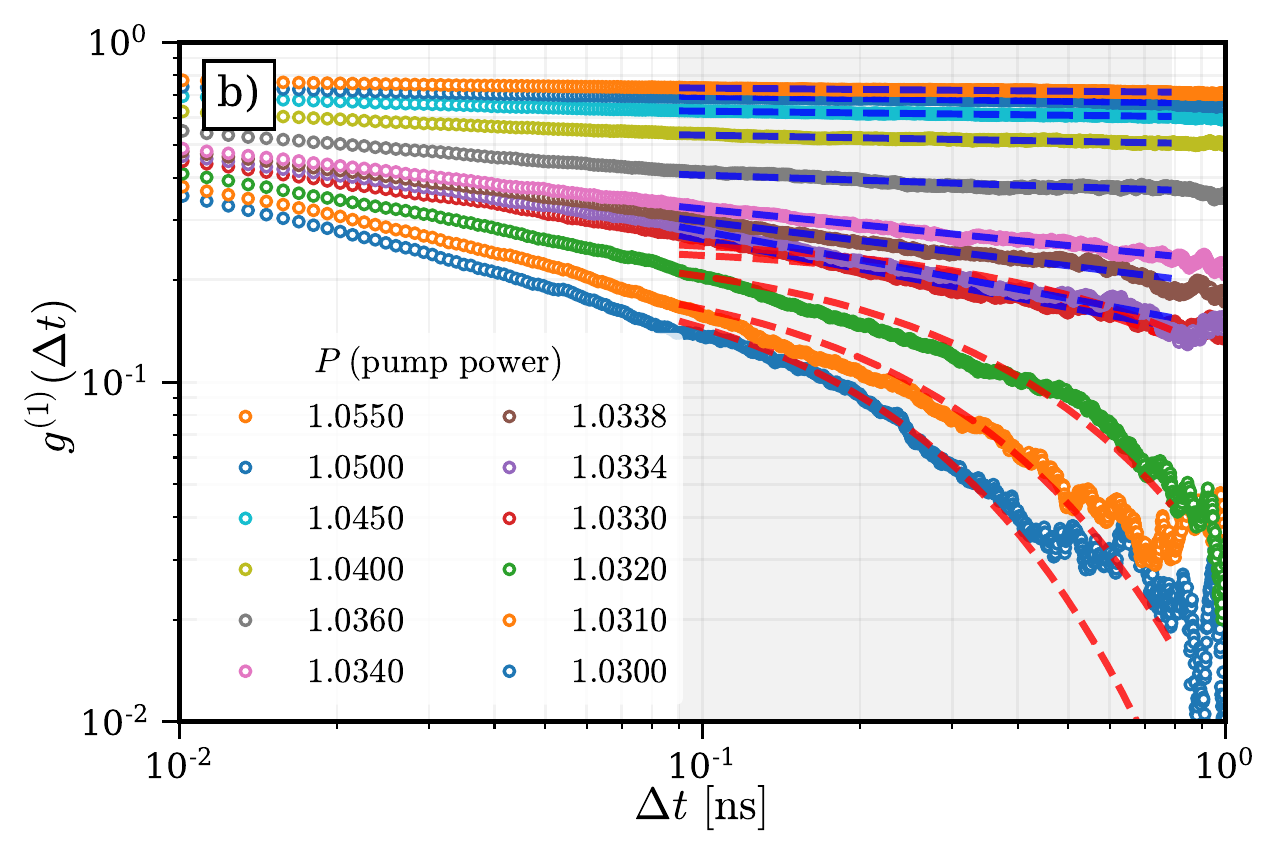}
	\includegraphics[width=.45\linewidth]{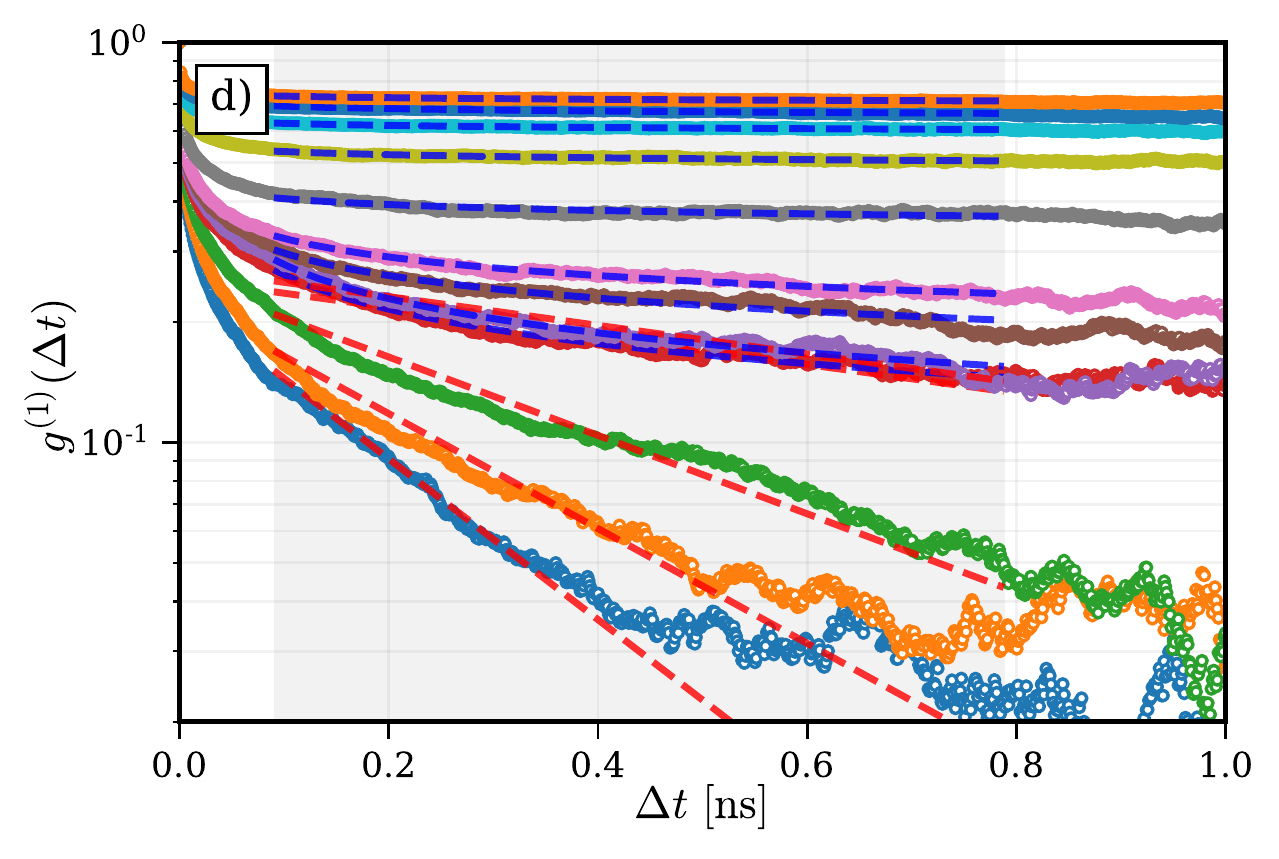}
	\caption{Spatial (top row) and temporal (bottom row) correlations, as shown in Fig.~(2) of the main text, both in log-log (left) and linear-log scale (right).
	}
	\label{fig:correlations_all_scales}
\end{figure*}
\begin{figure*}
	\centering
	\includegraphics[width=\linewidth]{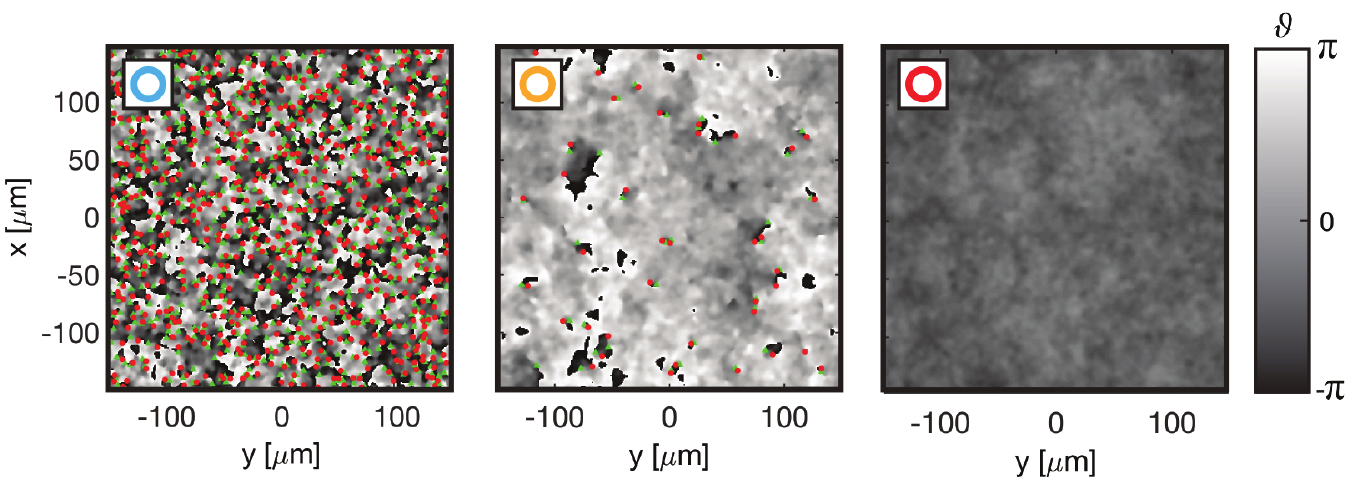}
	\caption{\textbf{Binding/unbinding crossover of topological defects.} 
		The three panels depict the phase ${\vartheta} = \arg{\{\psi\}}$ for (left) $P= 0.95$ with  $\left< N_\mathrm{v}\right> \sim 1.3 \times 10^{3}$, (centre) $P = 1.0338$ with $\left< N_\mathrm{v}\right> \sim 76$ and (right) $P = 1.06$ with $\left< N_\mathrm{v}\right> \sim 0$. Vortices and anti-vortices are reported as red dots and green triangles, respectively.
	}
	\label{fig:phases}
\end{figure*}
\begin{figure}
	\centering
	\includegraphics[width=.9\linewidth]{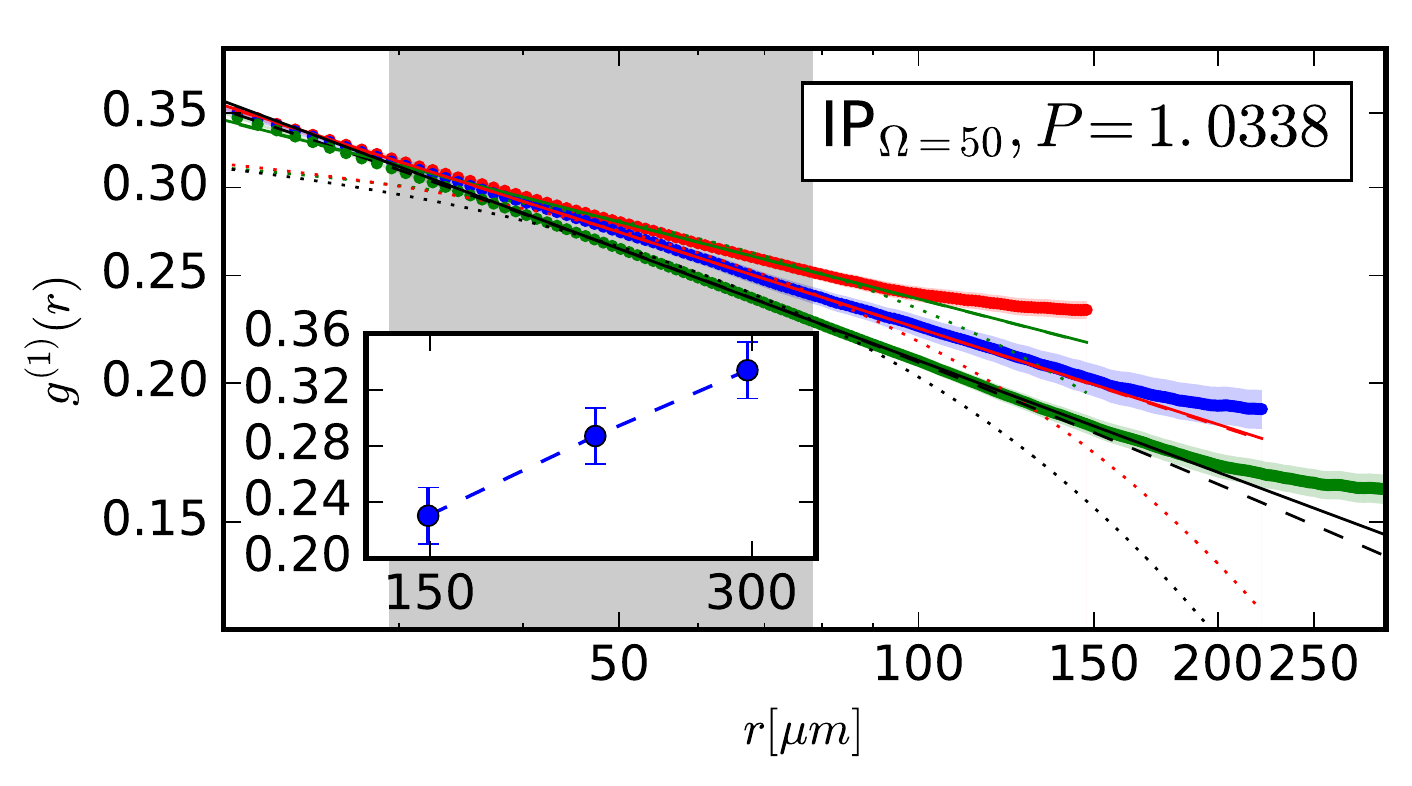}
	\includegraphics[width=.9\linewidth]{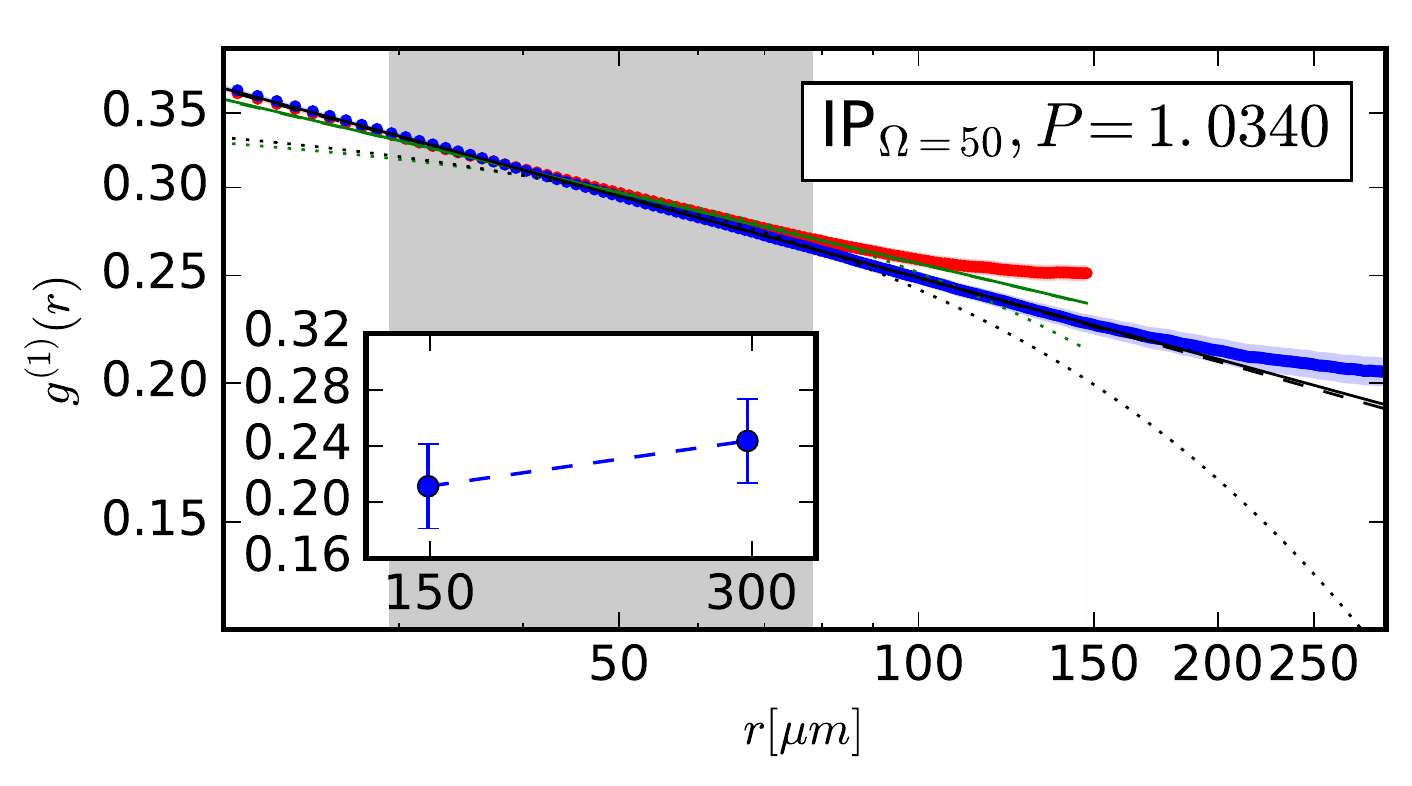}
	\caption{\textbf{Spatial correlations for larger system sizes.} Top panel: spatial correlations for $P = 1.0340$, immediately outside the critical region in the quasi-ordered phase, calculated for different lattice size $L=295.11 \mathrm{\mu m}$ (red points) and $L=444.42 \mathrm{\mu m}$ (blue points), for the case IP$_{\Omega = 50}$.
		Bottom panel: as top panel but for $P = 1.0338$ (middle), borderline case of the critical region, calculated for different lattice size $L=295.11 \mathrm{\mu m}$ (red points), $L=444.42 \mathrm{\mu m}$ (blue points) and $L=590.22 \mathrm{\mu m}$ (green points).
		Red points correspond to the spatial correlation for the ``standard'' box size, employed throughout the analysis of this chapter.
		In both panels power-law (solid line), stretched-exponential (dashed line) and exponential (dotted line) fits are reported.
		The exponent $\alpha$ as a function of the system size $L$ is reported in the insets of each panel. 
	}
	\label{fig:exponents}
\end{figure}

{\it Spatial first order correlation function.}
In Fig.~\ref{fig:correlations_all_scales}(a,c) we show the spatial first order correlation function $g^{(1)}(\Delta r)$ as a function of the pump strength $P$, calculated by means of \blue{Eq.~(8)} of the main text.
{To extract such a quantity, we first compute the two-point correlation function for each vertical and horizontal array of a single realisation of the wave function $\psi(\mathrm{\textbf{r}})$; we then average and normalize before additionally averaging over the $\mathcal{N}$ realizations, and computing corresponding error bars. }
We have moreover tested our results for \textit{i)} convergence in lattice size, \textit{ii)} convergence in stochastic numerical realisations \textit{iii)} convergence in computational method. More details are reported in the Supplemental Materials of Ref.~\cite{comaron2018dynamical}.

{\it Temporal first order correlation function.}
Fig.~\ref{fig:correlations_all_scales}(b,d) shows the temporal first order correlation function $g^{(1)}(\Delta t)$, calculated by means of Eq.~\blue{(9)} of the main text.
First, we make sure we reach the steady state at a given pump power $P$ and we then continue running the simulation for a long enough time (in our case we push the system evolution up to $ \Delta t ^{max}\sim 1 \mathrm{ns}$). 
In Eq.~\blue{(9)} of the main text, $\Delta t$ correspond to the relative time between the two central points of the two fields considered. 
In the vicinity of the critical point, we expect strong density and phase fluctuations. Therefore, a solid averaging procedure is crucial to achieve smooth and clean results for the correlator curves; this is important for extracting exponents as precise as possible.
In order to reach this, we average over realisations $\mathcal{N}$, but also over different times $t_i > t_0$ with same temporal windows, checking that different $t_i$ are sufficiently far apart so that any memory/correlations of the system are lost.
Therefore, a number of numerical results have been carried out. Specifically, we test: \textit{i)} 
convergence over averaging upon different initial temporal points $t_0$ chosen to be sufficiently far apart in time (see aforementioned discussion),
\textit{ii)} convergence in the choice of the initial $t_0$ (also in order to confirm that steady-state is reached for the minimum $t_0$ chosen), \textit{iii)} convergence in the number of stochastic paths $\mathcal{N}$.

We now discuss the procedure for evaluating the exponents and their errorbars. 
{Fig.~\blue{4}(a) of the main text illustrates how the critical region is inferred by the decay of the two-point correlators; here the Root Mean Deviation (RMSD) of the residuals of the fits are calculated as 
	\begin{equation}
		\mathrm{RMSD} = \sqrt{\frac{1}{n} \sum_{n}\left( d_i - f_i \right)^2},
	\end{equation}
	where $d_i$ and $f_i$ are the data points and the fit points at a position $j$ in space within the fitting window, with $j  = 1, \dots, n$.}
For both spatial and temporal correlators, the exponents plotted in Fig.~\blue{4}(b,c) of the main text are extracted by choosing a sensible choice of the spatial (temporal) window of the fits (gray shadows in Fig.~\blue{3}).
These are chosen so to avoid discretisation effects at small scales (times) and finite-size effects at large scales (late-time fluctuations given by the stochastic nature of the problem).
We check the robustness of such a fitting-window choice, and extract relative errorbars for the power-law exponents (e.g. in Fig.~\blue{4}(b,c) of the main text), by quantifying how the spatial (temporal) exponents change by varying the fitting window limits by $20 \%$ of their default values.

\section{Numerical counting of topological defects}
The number of topological defects is evaluated from the phase gradients around closed
paths of each grid point. First, a  Gaussian (low-pass) filter is applied to the wave function, in order to remove all the high frequency noise components. This removes all noise with wavelength (in pixels) smaller than, or of the order of, the standard deviation of the filter’s Gaussian kernel.
We proceed by extracting the phase, $\vartheta$, and gradient of the phase, $\nabla \vartheta = v$, using finite difference methods.
Finally, a vortex is identified (together with its location and charge) in a specific grid point when the circulation $\Gamma = \int_C v d{r} \gtrsim 2 \pi$, around a close path $C$ of double the size of the vortex healing length.

\section{Binding/unbinding crossover of topological defects}

In Fig.~\blue{2} of the main text, the average number of topological defects $\left< N_\mathrm{v} \right>$ at steady-state are plotted as a thick red line.
The condensate phase ${\vartheta} = \arg{\{\psi\}}$  of three exemplary configurations of vortices, below, in the vicinity of and above the BKT threshold are illustrated in Fig.~\ref{fig:phases}.
The left panel corresponds to a low-pump disordered phase (left), characterized by a high density of free vortices which are free to proliferate. 
{In spite of} the high density of topological defects $d_\mathrm{v} = N_\mathrm{v}/(L_x L_y)$ at this stage, we {have ensured} 
that {the vortex density remains} always smaller that the {polariton density} $\left < | \psi (t)|^2 \right >$, in order {for our physical description of vortices to be meaningful. Already for $P \sim 0.9$,} we find that {$d_\mathrm{v} \sim 0.01 \mathrm{\mu m^{-2}} $} and $d_\mathrm{v} / \left < | \psi (t)|^2 \right > \sim 0.01 $.
At the threshold point, $P \sim P_\mathrm{BKT}$, (centre) the process of vortices pairing starts to take place and is depicted. At high pump values $P \gg P_\mathrm{BKT}$, in the ordered phase (right), the system is left free of defects.

\section{Numerical extraction of the critical momentum}
In \blue{Fig.~5} of the main text we discuss three exemplary cases of excitation spectrum calculated from the spatio-temporal Fourier Transform (FT) of $|\psi{(\textbf{r}, t)}|^2$ across the PT.
For the latter case $P=1.06$ [panel $\textrm{(iv)}$] the critical momentum $k_c(P=1.06) = 1.26 \times 10^{-2} \mathrm{\mu m^{-1}}$ (calculated by means of \blue{Eq.~(5)} of the main text) is found on the order of the momentum discretization $\Delta k = \pi/L=1.06 \times 10^{-2} \mathrm{\mu m^{-1}}$ of the numerical simulation.
In order to extract numerically the value of $k_c$, a system $100 \times$ larger in size (but with same number of grid points)
is simulated and plotted in the inset of \blue{Fig.~{5}(iv)}. 
Finally, with such a choice of parameter, we can access a system with ($100 \times$) smaller $\Delta k$. 
We note that while with such a spatial discretization the model is no more able to capture the microscopic physics of the system, the momentum discretization $\Delta k$ is sufficiently small to measure a critical momentum  $k_c^\textrm{num} = 1.0(5) \times 10^{-2} \mathrm{\mu m^{-1}}$, in good agreement with the analytically predicted curve [\blue{Eq.~(6)} of the main text)]. 
In future works, it would be of a great interest simulating smaller numerical spatial spacing in such a large systems, so to be able to quantitatively study the effects of the microscale physics (accessible by means of a smaller numerical spatial spacing) on the value of the numerical critical momentum $k_c^\mathrm{num}$.

\begin{figure}[t]
	\centering
	\includegraphics[width=\linewidth]{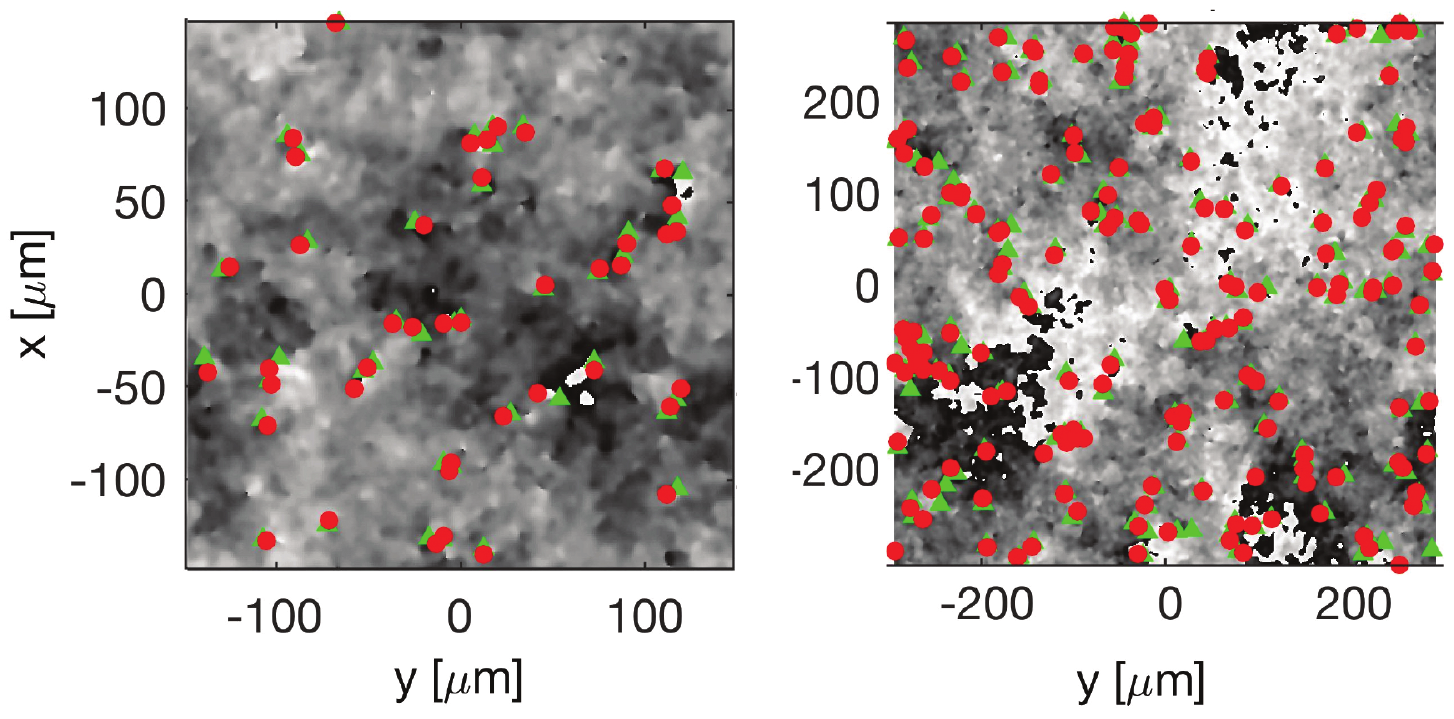}
	\caption{	
		\textbf{Vortex-antivortex pairs for the borderline case.}
		Filtered phase with detected vortices (red dots) and anti-vortices (green dots) are displayed for the case $P = 1.0338$, for $L=295.11 \mathrm{\mu m}$ (left) and $L=444.42 \mathrm{\mu m}$ (right), for the case IP$_{\Omega = 50}$. We note that only vortex-antivortex pairs are present in both configurations, suggesting the system is in a quasi-ordered phase. {Colormap as in Fig.~\ref{fig:phases}}.
	}
	\label{fig:exponents_vortices}
\end{figure}

\section{Long-range behaviour at the critical point}
\label{sec:g1_critical}

\begin{figure}
	\centering
	\includegraphics[width=\linewidth]{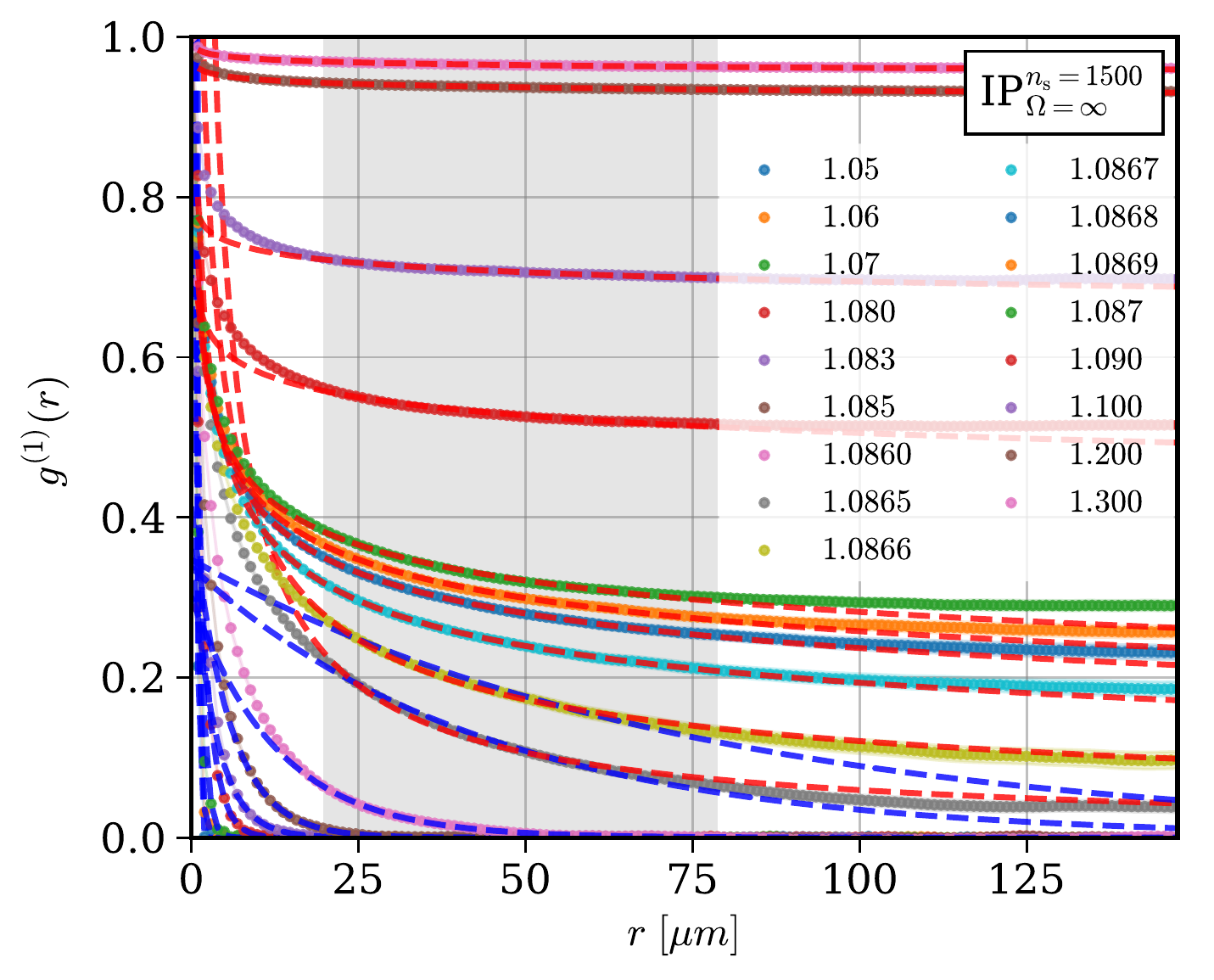}
	\includegraphics[width=\linewidth]{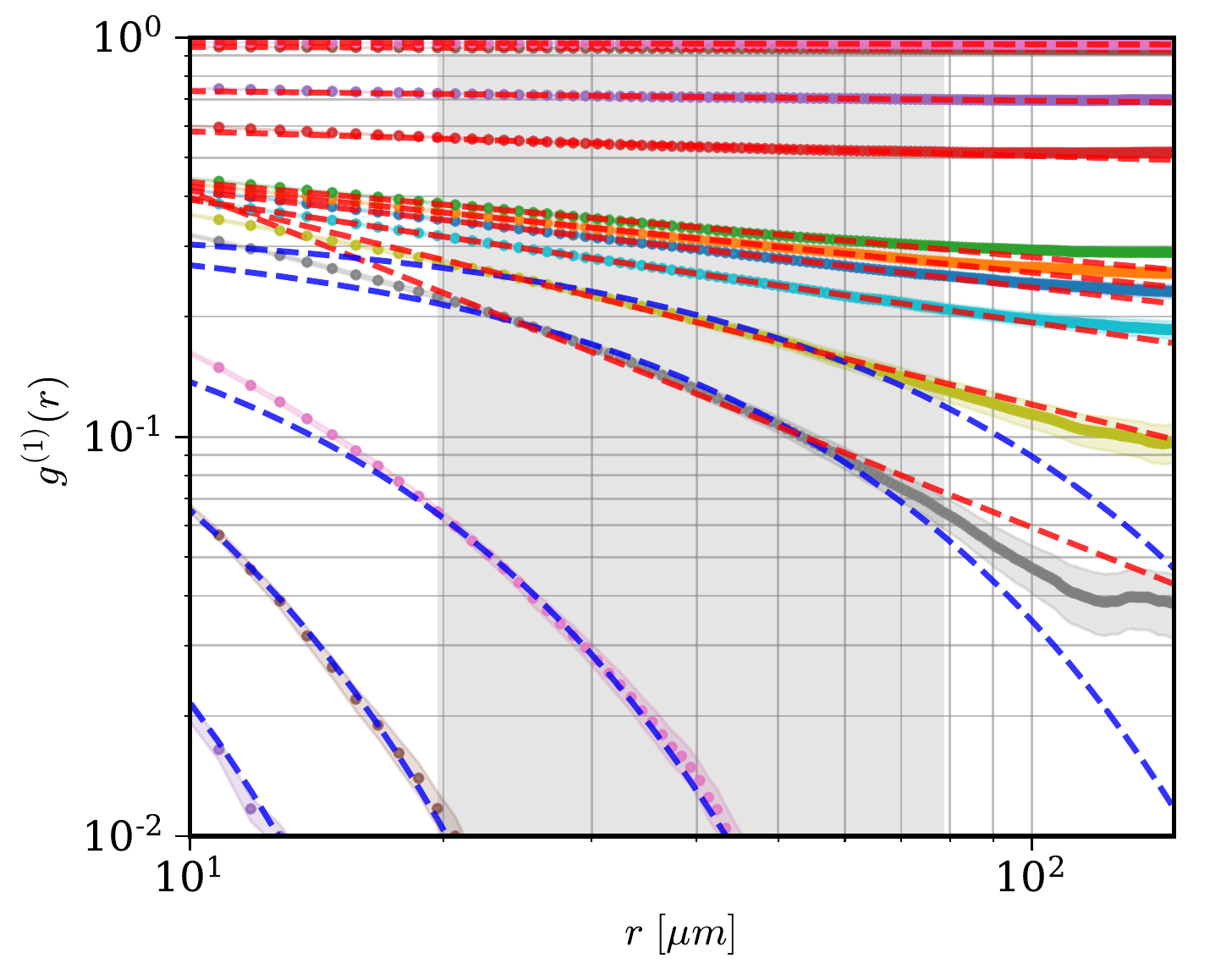}
	\caption{{\textbf{Exponential and algebraic decay crossover of spatial correlations for IP$_{\Omega = \infty}^{n_\mathrm{s}=1500}$.}} 
		As Fig.~\blue{3}(a) of the main text, but for the case IP$_{\Omega = \infty}^{n_\mathrm{s}=1500}$, in log-linear (top) and log-log (bottom) scales. The values of the pump powers $P$ are reported in the legend.
	}
	\label{fig:g1_omINF}
\end{figure}

In this section we describe the behaviour of the correlation function in the vicinity of a critical point in the (quasi)ordered phase; 
we discuss presence of possible finite-size effects and the critical values of the algebraic exponent around the BKT threshold.
As in the main text, we focus on the case IP$_{\Omega = 50}$ for which we identify the upper bound of the critical region to be located at $P=1.0338$, when the first (spatial and temporal) correlator shows ``clear'' power-law decay. 
Differently from the IP$_{\Omega = \infty}^{n_\mathrm{s}=1500}$ case, for the system size and parameter set considered in the main text analysis, the maximum exponent extracted does not exceed the equilibrium upper-bound limit, i.e. $\alpha_s< 0.25$.

We then proceed by investigating finite-size scaling properties around this limit.
Fig.~\ref{fig:exponents} shows comparison of spatial correlation functions $g^{(1)}(\Delta r)$ calculated for different box sizes, for $P = 1.0338$ (top), the borderline case of the ordered regime, and for a point just outside the critical region $P = 1.034$ (bottom).
Red points correspond to the correlations computed with a lattice size $L = 295.11 \mu m$ and $N=301$ points  (as in main text).
Blue and green points (when shown) correspond to $L = 444.42 \mu m$ and $L = 590.22 \mu m$, i.e. to a box size enlarged by $50 \%$  and $100 \%$, respectively.
The different curves are then fitted with exponential and power-law functions, as in the analysis performed in the main text. 
We note that while for the case $P = 1.034$, the curves are converged within the fitting area (grey faint region), confirming the power-law decay for larger boxes, the case $P = 1.0338$ exhibits a further decrease of the slope as the system size increases, suggesting that the correlators are indeed not converged, with finite size effects being still relevant and responsible for the changing of the curve shapes.
It is worth noting that such a behaviour is consistent with observations within the OPO framework \cite{dagvadorj2015nonequilibrium}.

However, we note that for the limit case $P = 1.0338$ the spatial correlators for experimentally-relevant system sizes still exhibit an algebraic decay (solid lines) in favour of an exponential behaviour (dashed curves).
Thus, we proceed evaluating the exponent $\alpha$ extracted from the power-law fit for the case $P=1.0338$.
In the insets of Fig.~\ref{fig:exponents}, we plot the exponent $\alpha$ as a function of the box size for each case.
We find that the default box size shows $\alpha = 0.23$ (just below the equilibrium threshold) while, for larger boxes $\alpha$ exceeds the limit of $1/4$, with $\alpha = 0.28$ and $\alpha=0.33$ for $L = 444.42 \mu m$ and $L = 590.22 \mu m$, respectively.
To consolidate our conclusions on the BKT-like nature of the system in this limit case, we investigate its topological defects configuration. A quasi-ordered regime is indeed confirmed by the presence of \textit{only} vortex pairs in the system, as illustrated in Fig.~\ref{fig:exponents_vortices} for one of the many realizations, for both smaller and larger boxes.

Concluding, in this section we demonstrated that in the vicinity of the critical point, within experimentally-relevant sizes, the system manifests an exponent $\alpha > 1/4$, therefore suggesting its nature to be of the ``non-equilibrium'' type, in agreement with previous works \cite{nitsche2014algebraic,dagvadorj2015nonequilibrium}.

\begin{figure}[]
	\centering
	\includegraphics[width=\linewidth]{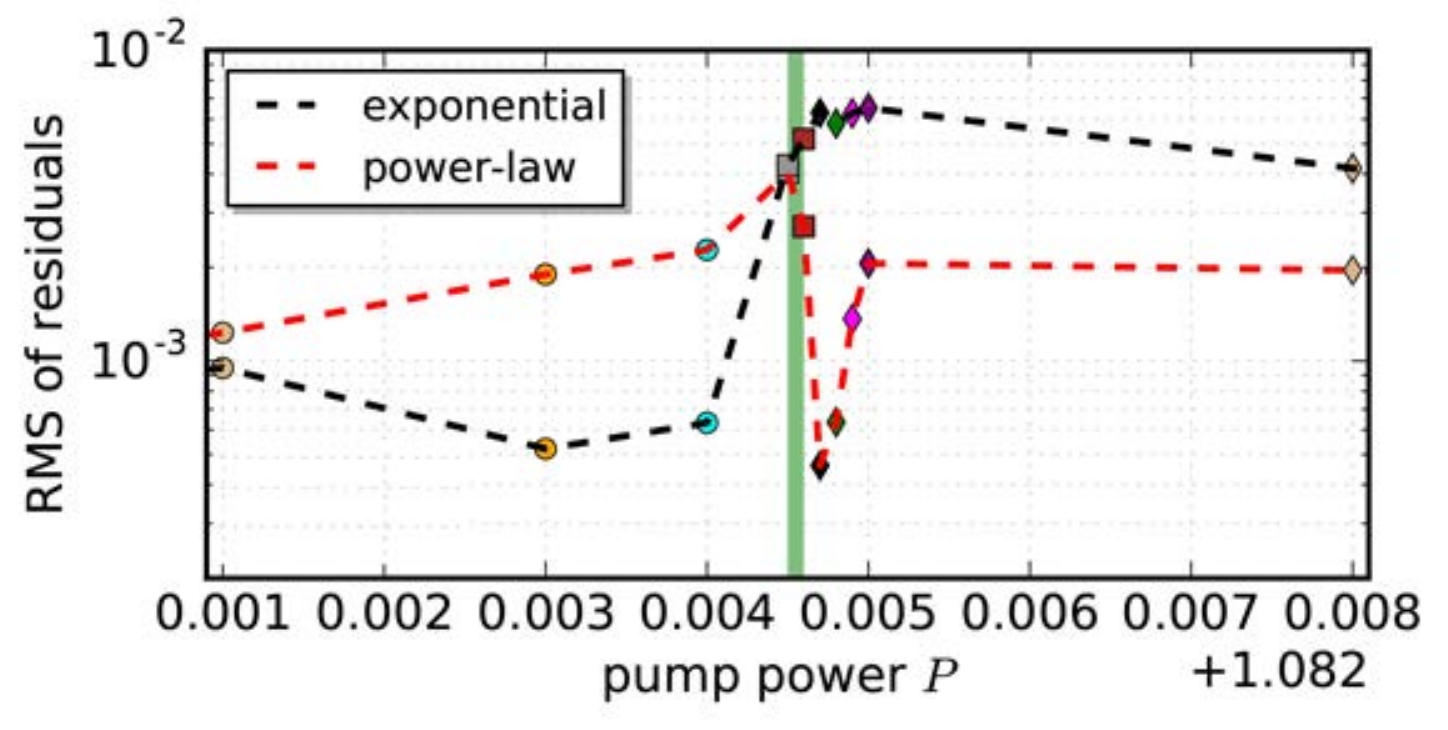}
	\caption{As Fig.~\blue{4}(a) of the main text, but for the case IP$_{\Omega = \infty}^{n_\mathrm{s}=1500}$} 
	\label{fig:exponents_bottom_omINF}
\end{figure} 

\section{Spatial coherence for the frequency-independent pump case IP$_{\Omega = \infty}^{n_\mathrm{s}=1500}$}
\label{sec:appendix_A}

In this section we present the fitting procedure and analysis for the frequency-independent pump case IP$_{\Omega = \infty}^{n_\mathrm{s}=1500}$.
Similarly to Fig.~\blue{3}(a) of the main text for the frequency-selective case, Fig.~\ref{fig:g1_omINF} shows the exponential-to-algebraic long-range decay crossover of the spatial  and temporal correlation functions for increasing pump power $P$.
Thick dashed red (blue) curves correspond to exponential (power-law) fitting, from which values of correlation length $\xi$ and power-law exponents $\alpha_\mathrm{s}$  are extracted and plotted in the inset of Fig.~\blue{4}(c).
For each curve, we superimpose only the appropriate fit, over the chosen fitting window (plotted as gray shadow); we show both fits for curves which lie in the critical region.
From this analysis, we find that the critical region lower and upper boundaries correspond to the values $P = 1.0865$ and $P = 1.0866$, respectively.
In Fig.~\ref{fig:exponents_bottom_omINF} we plot the RMDS of the fits.


\begin{thebibliography}{10}
	\expandafter\ifx\csname url\endcsname\relax\def\url#1{\texttt{#1}}\fi
	
	\bibitem{pitaevskii2003bose}
	\Name{Pitaevskii L. \and Stringari S.} \Book{Bose-Einstein Condensation}
	(Clarendon Press) 2003.
	
	\bibitem{MerminWagner1966}
	\Name{Mermin N.~D. \and Wagner H.} \REVIEW{Phys. Rev. Lett.}{17}{1966}{1133}.
	
	\bibitem{Berezinskii1971}
	\Name{Berezinskii V.~L.} \REVIEW{Sov. Phys. JETP}{32}{1971}{493}.
	
	\bibitem{Berezinskii1973}
	\Name{Kosterlitz J.~M. \and Thouless D.~J.} \REVIEW{Journal of Physics C: Solid
		State Physics}{6}{1973}{1181}.
	
	\bibitem{carusotto2013quantum}
	\Name{Carusotto I. \and Ciuti C.} \REVIEW{Reviews of Modern
		Physics}{85}{2013}{299}.
	
	\bibitem{proukakis_snoke_littlewood_2017}
	\Name{Proukakis N.~P., Snoke D.~W. \and Littlewood P.~B.} \Book{(Eds.)
		Universal Themes of Bose-Einstein Condensation} (Cambridge University Press)
	2017.
	
	\bibitem{hohenberg1977theory}
	\Name{Hohenberg P.~C. \and Halperin B.~I.} \REVIEW{Reviews of Modern
		Physics}{49}{1977}{435}.
	
	\bibitem{kasprzak2006bose}
	\Name{Kasprzak J., Richard M., Kundermann S., Baas A., Jeambrun P., Keeling J.,
		Marchetti F., Szyma{\'n}ska M., Andre R., Staehli J. \etal}
	\REVIEW{Nature}{443}{2006}{409}.
	
	\bibitem{caputo2018}
	\Name{Caputo D., Ballarini D., Dagvadorj G., S{\'a}nchez~Mu{\~n}oz C.,
		De~Giorgi M., Dominici L., West K., Pfeiffer L.~N., Gigli G., Laussy F.~P.,
		Szyma{\'n}ska M.~H. \and Sanvitto D.} \REVIEW{Nature Materials}{17}{2017}{145
		EP }.
	
	\bibitem{stepanov2019dispersion}
	\Name{Stepanov P., Amelio I., Rousset J.-G., Bloch J., Lema{\^\i}tre A., Amo
		A., Minguzzi A., Carusotto I. \and Richard M.} \REVIEW{Nature
		communications}{10}{2019}{1}.
	
	\bibitem{ballarini2020directional}
	\Name{Ballarini D., Caputo D., Dagvadorj G., Juggins R., De~Giorgi M., Dominici
		L., West K., Pfeiffer L.~N., Gigli G., Szyma{\'n}ska M.~H. \etal}
	\REVIEW{Nature communications}{11}{2020}{1}.
	
	\bibitem{proukakis2013quantum}
	\Name{Proukakis N.~P., Gardiner S.~A., Davis M.~J. \and Szyma{\'n}ska M.~H.}
	\Book{(Eds.) Quantum Gases: Finite Temperature and Non-Equilibrium Dynamics}
	Vol.~1 (World Scientific) 2013.
	
	\bibitem{Gladilin2019}
	\Name{Gladilin V.~N. \and Wouters M.} \REVIEW{Phys. Rev. B}{100}{2019}{214506}.
	
	\bibitem{Mei2019}
	\Name{Mei Q., Ji K. \and Wouters M.} \REVIEW{arxiv:2002.01806}{}{2019}{}.
	
	\bibitem{Richard2005}
	\Name{Richard M., Kasprzak J., Romestain R., Andr\'e R. \and Dang L.~S.}
	\REVIEW{Phys. Rev. Lett.}{94}{2005}{187401}.
	
	\bibitem{Wouters2008}
	\Name{Wouters M., Carusotto I. \and Ciuti C.} \REVIEW{Phys. Rev.
		B}{77}{2008}{115340}.
	
	\bibitem{szymanska2006nonequilibrium}
	\Name{Szyma{\'n}ska M., Keeling J. \and Littlewood P.} \REVIEW{Physical review
		letters}{96}{2006}{230602}.
	
	\bibitem{altman2015twodimensional}
	\Name{Altman E., Sieberer L.~M., Chen L., Diehl S. \and Toner J.} \REVIEW{Phys.
		Rev. X}{5}{2015}{011017}.
	
	\bibitem{dagvadorj2015nonequilibrium}
	\Name{Dagvadorj G., Fellows J., Matyja{\'s}kiewicz S., Marchetti F., Carusotto
		I. \and Szyma{\'n}ska M.} \REVIEW{Physical Review X}{5}{2015}{041028}.
	
	\bibitem{Roumpos6467}
	\Name{Roumpos G., Lohse M., Nitsche W.~H., Keeling J., Szyma{\'n}ska M.~H.,
		Littlewood P.~B., L{\"o}ffler A., H{\"o}fling S., Worschech L., Forchel A.
		\and Yamamoto Y.} \REVIEW{PNAS}{109}{2012}{6467}.
	
	\bibitem{szymanska2007mean}
	\Name{Szyma{\'n}ska M., Keeling J. \and Littlewood P.} \REVIEW{Physical Review
		B}{75}{2007}{195331}.
	
	\bibitem{Krizhanovskii2006}
	\Name{Krizhanovskii D.~N., Sanvitto D., Love A. P.~D., Skolnick M.~S.,
		Whittaker D.~M. \and Roberts J.~S.} \REVIEW{Phys. Rev.
		Lett.}{97}{2006}{097402}.
	
	\bibitem{Love2008}
	\Name{Love A. P.~D., Krizhanovskii D.~N., Whittaker D.~M., Bouchekioua R.,
		Sanvitto D., Rizeiqi S.~A., Bradley R., Skolnick M.~S., Eastham P.~R.,
		Andr\'e R. \and Dang L.~S.} \REVIEW{Phys. Rev. Lett.}{101}{2008}{067404}.
	
	\bibitem{Kim2016}
	\Name{Kim S., Zhang B., Wang Z., Fischer J., Brodbeck S., Kamp M., Schneider
		C., H\"ofling S. \and Deng H.} \REVIEW{Phys. Rev. X}{6}{2016}{011026}.
	
	\bibitem{Askitopoulos2019}
	\Name{Askitopoulos A., Pickup L., Alyatkin S., Zasedatelev A., Lagoudakis K.,
		Langbein W. \and Lagoudakis P.~G.} \REVIEW{arXiv:1911.08981}{}{2019}{}.
	
	\bibitem{AmelioCarusotto2020}
	\Name{Amelio I. \and Carusotto I.} \REVIEW{Phys. Rev. X}{10}{2020}{041060}.
	
	\bibitem{WC2007}
	\Name{Wouters M. \and Carusotto I.} \REVIEW{Phys. Rev.
		Lett.}{99}{2007}{140402}.
	
	\bibitem{woutersLiew2010}
	\Name{Wouters M., Liew T. C.~H. \and Savona V.} \REVIEW{Phys. Rev.
		B}{82}{2010}{245315}.
	
	\bibitem{chiocchetta2013}
	\Name{Chiocchetta A. \and Carusotto I.} \REVIEW{EPL}{102}{2013}{67007}.
	
	\bibitem{comaron2018dynamical}
	\Name{Comaron P., Dagvadorj G., Zamora A., Carusotto I., Proukakis N.~P. \and
		Szyma\ifmmode~\acute{n}\else \'{n}\fi{}ska M.~H.} \REVIEW{Phys. Rev.
		Lett.}{121}{2018}{095302}.
	
	\bibitem{WoutersCarusotto2010}
	\Name{Wouters M. \and Carusotto I.} \REVIEW{Phys. Rev.
		Lett.}{105}{2010}{020602}.
	
	\bibitem{SM}
	See [link] for numerical details and methods.
	
	\bibitem{zamora2020}
	\Name{Zamora A., Dagvadorj G., Comaron P., Carusotto I., Proukakis N.~P. \and
		Szyma\ifmmode~\acute{n}\else \'{n}\fi{}ska M.~H.} \REVIEW{Phys. Rev.
		Lett.}{125}{2020}{095301}.
	
	\bibitem{Comaron2019}
	\Name{Comaron P., Larcher F., Dalfovo F. \and Proukakis N.~P.} \REVIEW{Phys.
		Rev. A}{100}{2019}{033618}.
	
	\bibitem{keeling2010keldysh}
	\Name{Keeling J., Szyma{\'n}ska M.~H. \and Littlewood P.~B.} \Book{Keldysh
		Green’s function approach to coherence in a non-equilibrium steady state}
	(Springer) 2010 pp. 293--329.
	
\end{thebibliography}

\begin{thebibliography}{1}
	\expandafter\ifx\csname url\endcsname\relax\def\url#1{\texttt{#1}}\fi
	
	\bibitem{Sinatra2008}
	\Name{Sinatra A. \and Castin Y.} \REVIEW{Phys. Rev. A}{78}{2008}{053615}.
	
	\bibitem{carusotto2013quantum}
	\Name{Carusotto I. \and Ciuti C.} \REVIEW{Reviews of Modern
		Physics}{85}{2013}{299}.
	
	\bibitem{comaron2018dynamical}
	\Name{Comaron P., Dagvadorj G., Zamora A., Carusotto I., Proukakis N.~P. \and
		Szyma\ifmmode~\acute{n}\else \'{n}\fi{}ska M.~H.} \REVIEW{Phys. Rev.
		Lett.}{121}{2018}{095302}.
	
	\bibitem{dennis2013xmds2}
	\Name{Graham R., Hope J. \and Johnsson M.} \REVIEW{Comp. Phys.
		Commun.}{184}{2013}{201}.
	
	\bibitem{nitsche2014algebraic}
	\Name{Nitsche W. \etal} \REVIEW{Phys. Rev. B}{90}{2014}{205430}.
	
	\bibitem{zamora2020}
	\Name{Zamora A., Dagvadorj G., Comaron P., Carusotto I., Proukakis N.~P. \and
		Szyma\ifmmode~\acute{n}\else \'{n}\fi{}ska M.~H.} \REVIEW{Phys. Rev.
		Lett.}{125}{2020}{095301}.
	
	\bibitem{dagvadorj2015nonequilibrium}
	\Name{Dagvadorj G., Fellows J., Matyja{\'s}kiewicz S., Marchetti F., Carusotto
		I. \and Szyma{\'n}ska M.} \REVIEW{Physical Review X}{5}{2015}{041028}.
	
\end{thebibliography}
\end{document}